\documentclass[10pt]{article}
\usepackage{amsfonts}

\input epsf.tex

\def\str{{\rm STr \;}}
\def\({\left(}
\def\){\right)}
\def\th{\theta}         
\def\ga{\gamma}         

\def\al{\alpha}
\def\ep{\epsilon}
\def\la{\lambda}        
         \def\De{\Delta}
\def\om{\omega}         
\def\sig{\sigma}        \def\Sig{\Sigma}
\def\vphi{\varphi}
              \def\CC{{\cal C}}
\def\CD{{\cal D}}

\def\CS{{\cal S}}              
              
       \def\CZ{{\cal Z}}
\begin{document}

\title{Localization of quasiparticles in a disordered vortex}

\author{
R.~Bundschuh\footnote{Current address: 
Department of Physics, UCSD, La Jolla, CA  92093-0319, U.S.A.}
, C.~Cassanello, 
D.~Serban\footnote{Current address: Service de Physique Th\'eorique, 
CE-Saclay, F-91191 Gif-Sur-Yvette, France.} 
, and M.R.~Zirnbauer\\ {}\\
{\it Institut f\"ur Theoretische Physik}\\
{\it Universit\"at zu K\"oln, Z\"ulpicher Str. 77, D-50937, K\"oln, Germany}}
\date{}
\maketitle

%%%%%%%%%%%%%%%%%%%%%%%%%%%%%%%%%%%%%%%% 
\newif\ifbbB\bbBfalse                %%% 
%%%       BLACKBOARD BOLD FONT       %%% 
%%% Comment-out the next line if you %%% 
%%% do NOT have the blackboard bold  %%% 
%%% (mssb) fonts:                    %%% 
\bbBtrue                             %%% 
%%%%%%%%%%%%%%%%%%%%%%%%%%%%%%%%%%%%%%%% 
\ifbbB 
  \message{If you do not have msbm (blackboard bold) fonts,}
  \message{change the option at the top of the text file.}
  \font\blackboard=msbm10 
  \font\blackboards=msbm7
  \font\blackboardss=msbm5 
  \newfam\black \textfont\black=\blackboard
  \scriptfont\black=\blackboards 
  \scriptscriptfont\black=\blackboardss
  \def\Bbb#1{{\fam\black\relax#1}} 
\else 
  \def\Bbb{\bf} 
\fi

\begin{abstract}
  We study the diffusive motion of low-energy normal quasiparticles
  along the core of a single vortex in a dirty, type-II, s-wave
  superconductor.  The physics of this system is argued to be
  described by a one-dimensional supersymmetric nonlinear $\sigma$
  model, which differs from the $\sigma$ models known for disordered
  metallic wires.  For an isolated vortex and quasiparticle energies
  less than the Thouless energy $E_{\rm Th}$, we recover the spectral
  correlations that are predicted by random matrix theory for the
  universality class $C$.  We then consider the transport problem of
  transmission of quasiparticles through a vortex connected to
  particle reservoirs at both ends.  The transmittance at zero energy
  exhibits a weak localization correction reminiscent of
  quasi-one-dimensional metallic systems with symmetry index $\beta =
  1$.  Weak localization disappears with increasing energy over a
  scale set by $E_{\rm Th}$.  This crossover should be observable in
  measurements of the longitudinal heat conductivity of an ensemble of
  vortices under mesoscopic conditions.  In the regime of strong
  localization, the localization length is shown to decrease by a
  factor of 8 as the quasiparticle energy goes to zero.
\end{abstract}

\bigskip{PACS numbers: 74.40.+k, 74.25.Fy, 72.80.Ng, 73.23.-b}

\section{Introduction}

Disordered single-electron systems fall into several universality
classes distinguished by their symmetries under time reversal and
rotation of the electron spin.  In mesoscopic physics three such
classes are well known, namely those anticipated by Dyson's 1962
classification \cite{dyson} of many-body systems with complex
interactions.  They are called the classes with orthogonal, unitary,
or symplectic symmetry. In the ergodic regime, where the relevant time
scale is much longer than the time for an electron to diffuse across
the system, each universality class is described by a Wigner-Dyson
random matrix ensemble with symmetry index $\beta = 1, 2, 4$,
respectively.

In the present article we consider a fascinating variant of the theme
of disordered electron systems: a single vortex in a disordered
type-II conventional s-wave superconductor.  A simple picture would
model the vortex as a tube-shaped metallic region inside the
superconductor.  Indeed, a vortex supports quasiparticle excitations,
which have normal characteristics and are bound to the vortex core by
the pairing gap of the superconductor.  It is then natural to ask how
the quasiparticles of a disordered vortex fit into the traditional
classification scheme.

The first and important message is that the low-energy quasiparticles
exceed the traditional scheme and form a separate universality class.
This assertion can be motivated as follows.  Dyson's classification
assumes the absence of symmetries of the particle-hole type.  When
this assumption is relaxed, seven more universality classes enter.
Three of these are realized for a massless Dirac particle moving in a
random gauge field \cite{verbaarschot}, and another four have recently
been identified \cite{az_ns} in metallic systems in contact with a
superconductor.  Common to all of these classes is an invariance of
the energy (or Dirac) spectrum under inversion $E \mapsto -E$.  In the
limit $E \to \infty$, the inversion symmetry becomes irrelevant and
conventional Wigner-Dyson statistics is recovered, but for small
enough $E$ novel features appear.  In particular, the disorder average
of the density of states is nonstationary and yet universal.

A vortex is an example of a metallic system with particle-hole
symmetry.  (Note that this particle-hole symmetry is not a dynamical
symmetry but arises as a result of doubling the single-particle
Hilbert space \`a la Bogoliubov-deGennes, in order to incorporate
pairing correlations within the formalism of first quantization.)  The
magnetic field admitted by the vortex breaks time reversal invariance
whereas, if magnetic impurities are absent, the spin of a
quasiparticle is conserved.  According to the classification scheme of
\cite{az_ns}, this means that a disordered vortex belongs to symmetry
class $C$.  Moreover, the supercurrent circulating around a vortex
causes the phase shift due to Andreev scattering to vanish on average,
so that by the reasoning of \cite{az_ns} the Gaussian random matrix
ensemble of type $C$ is expected to be appropriate at low excitation
energies.

These predictions have recently been applied to dissipative vortex
motion in a quick proposal (for pancake vortices) by Skvortsov and
Feigel'man \cite{feigelman}.  The basic picture is as follows.
Dissipation is due to the transfer of energy from the collective
motion to the internal degrees of freedom.  In the case of a vortex
the relevant internal degrees of freedom are those of the
quasiparticles.  The rate of energy transfer is largely determined by
the statistical properties of the quasiparticle wavefunctions and
energies.  Therefore, a modification of their statistics is expected
to modify the rate of dissipation.  Indeed, Skvortsov and Feigel'man
argued that the level statistics of type $C$ leads to a dissipative
flux-flow conductivity which is weakly anomalous.

Although random matrix theory in its universal form is a powerful and
successful concept, it is limited to the ergodic regime of times
larger than the diffusion time or, equivalently, energies smaller than
the Thouless energy.  At short times, or high energies, the diffusive
dynamics of a disordered electron system causes deviations from random
matrix statistics.  This is true for ordinary metals and must equally
be true for the quasiparticles of a vortex.  Clearly, a complete
theory of mesoscopic phenomena needs to encompass both regimes and
make the crossover between them theoretically accessible.  For the
case of ordinary metals, such a theory exists and is in good shape.
In its most elegant formulation, it is given by the diffusive
nonlinear $\sigma$ model pioneered by Wegner \cite{wegner} and
perfected by Efetov \cite{efetov}.  This is a nonlinear field theory
of interacting diffusion modes, which has random matrix theory for its
universal low-energy limit.

Is it possible to establish a similar description for the
quasiparticles of a disordered vortex?  As follows from the above
discussion, such a description is of relevance to the microscopic
theory of dissipative vortex dynamics.  It should also be directly
testable by experiment, by measuring the longitudinal heat
conductivity to probe the transport properties of the quasiparticles
at low energies or temperatures.  (Note that electrical conductance
measurements are ruled out because the vortex is short cut by the
superconducting condensate.)

It is clear that the construction of a field theory for the disordered
vortex, encompassing both the diffusive quasiparticle dynamics and its
ergodic (or random matrix) long time limit, is a more intricate
problem than deriving the nonlinear $\sigma$ model for disordered
metals.  This is because diffusion in a disordered superconductor is
already a nontrivial nonlinear process.  The relevant kinetic equation
was derived by Usadel \cite{usadel}, building on earlier work by
Eilenberger \cite{eilenberger}.  For the special geometry of a single
vortex, the solution of the Usadel equation was obtained in Refs.
\cite{usadelvortex}, \cite{kramer} to predict the crude structure of
the local density of states of the vortex and its thermodynamic
properties.

The Usadel equation is well established as a tool in the research area
of disordered superconductors and the proximity effect.  However,
because of its quasiclassical nature, it leaves out mesoscopic effects
that are due to quantum coherence, such as weak localization and
mesoscopic fluctuation phenomena.  The question then is whether one
can improve on the quasiclassical Usadel equation and incorporate the
relevant quantum effects which, in the long time limit, give rise to
random matrix features.  The answer is yes.  In an important recent
letter, Altland {\it et al.} \cite{ats} pointed out that the Usadel
equation can be interpreted as the saddle point equation for a certain
supersymmetric field theory of the type of a nonlinear $\sigma$ model.
(The general connection between quasiclassical kinetics and nonlinear
$\sigma$ models was first understood by Muzykantskii and Khmel'nitskii
\cite{mk}.)  The formalism of Altland {\it et al.} also shows how to go
further and include quantum interference effects, by computing the
fluctuations around the saddle point solution.  In the present paper
this formalism will be applied to the case of a disordered vortex.

A brief account of the contents of this paper is as follows.  Section
\ref{sec:model} defines the model and sets up the basic field
theoretic formalism, the main step being the introduction of a
composite field $Q$ by means of a Hubbard-Stratonovich transformation.
In Section \ref{sec:saddle}, two saddle point approximations are made
in sequence.  The first of these restricts $Q$ to a nonlinear field
space obeying the local constraint $Q^2(x) = 1$.  As usual, this
approximation is justified for {\it diffusive} systems, where the
disorder is at the same time (i) sufficiently strong to cause rapid
momentum relaxation (thereby leading to diffusive motion in coordinate
space) and (ii) weak enough in order for the system to retain metallic
characteristics.

The second saddle point approximation is made in the course of solving
for the dynamics of the $Q$ field {\it inside} the nonlinear space
$Q^2 = 1$.  To that end, we expand the full Lagrangian of the theory
to obtain the relevant ``low-energy'' Lagrangian, containing in
addition to the usual diffusive (or kinetic) term a linear coupling to
the superconducting order parameter, and the minimal coupling to the
magnetic field present in the vortex.  Such an expansion is valid in
the so-called ``dirty limit'', where the coherence length $\xi_0$
({\it i.e.} the length scale of variation of the pairing field)
exceeds the elastic mean-free path $\ell$.  The equation of motion
obtained by varying the resulting action functional turns out to be
the Usadel equation.  Its solution depends only on the scaling
variable $(\Delta_0 / D)^{1/2} r$, where $r$ is the distance from the
axis of the vortex, $\Delta_0$ is the magnitude of the pairing gap in
the bulk, and $D$ is the diffusion constant of the normal metal.  This
result follows on dimensional grounds and is in agreement with the
well-known fact \cite{BdG} that the coherence length gets reduced by a
factor $(\ell/\xi_0)^{1/2}$ in the dirty limit.

Section \ref{sec:onedim} explores the consequences of our key
observation, which is that the solution of the Usadel equation is {\it
  degenerate}.  More precisely, the field theory for $Q$ has an exact
global symmetry broken only by the energy distance from the Fermi
level.  (This symmetry is not present in traditional work based on the
Usadel equation.  We believe that its existence can readily be seen
only when the field theoretic formalism is used.)  Letting the
symmetry group act on the solution of the Usadel equation we get a
degenerate {\it manifold} of solutions.  The ``radius'' of the
manifold is maximal on the axis of the vortex and shrinks to zero with
increasing distance, over a scale given by $( D / \Delta_0)^{1/2} \sim
(\xi_0 \ell)^{1/2}$.  Thus, the symmetry becomes ineffective as we
move away from the vortex to the asymptotic region.  In this way the
symmetry, while being {\it global} or constant in space, manages to
perform the curious trick of having an effect that is {\it localized}
in space!

For an isolated vortex of length $L$, the statistics of the
quasiparticle levels at energies less than the Thouless energy $E_{\rm
  Th} \sim D / L^2$ is determined by the manifold swept out by the
symmetry group when acting on the solution of Usadel's equation.  By
integrating over it, we recover for the vortex the random matrix
prediction for the level statistics of systems in universality class
$C$.  For higher energies $(E > E_{\rm Th})$, or fixed energy and
increasing system size $L$, long wave length fluctuations inside the
manifold become important.  These are governed by a one-dimensional
field theory of the nonlinear $\sigma$ model type.  We identify the
field space as a Riemannian symmetric superspace of type $D{\rm III} |
C{\rm I}$.  The field is called $q$.

In Section \ref{sec:trans} we turn to the investigation of transport
properties of the quasiparticle excitations of the disordered vortex.
We imagine the two ends of the vortex (or of an ensemble of vortices,
to facilitate experiment) to be connected to infinite particle
reservoirs and study the problem of transmission of quasiparticles
along the axis of the vortex.  We work in a mesoscopic regime,
requiring low energy and temperature.  This problem is the natural
analog of calculating the disorder average of the conductance of a
mesoscopic metallic wire.  The specific observable we study is a
correlation function for the conserved probability current carried by
quasiparticles of energy $E$.  We call this observable the
``transmittance'', $\tau(E)$.  The coupling to the reservoirs is
modeled by open boundary conditions on the field $q$ in the standard
way.

We then carry out the usual field theoretic perturbation expansion up
to one-loop order.  To deal with the symmetry breaking due to finite
energy $E$, which causes crossover between universality classes, we
introduce a parametrization of the fields akin to that used by
Altland, Iida and Efetov \cite{ida} in their supersymmetric approach
to the crossover problem for small disordered electron systems with
orthogonal and unitary symmetry. The result we obtain is
      $$
      \tau = {\xi_{\rm loc} \over L} - {\rm Re} \left( 
        { \coth\sqrt{i\epsilon} \over \sqrt{i\epsilon} } \right) + 
        {\cal O}(L/\xi_{\rm loc}) \;,
      $$
where $\xi_{\rm loc}$ has the meaning of a localization length, and
$\epsilon$ is the energy measured in units of (roughly speaking) the
Thouless energy.  For quasiparticles of zero energy, the result
simplifies to $\tau = (\xi_{\rm loc}/L) - (1/3) + ...$, which is
similar to the conductance of quasi-one-dimensional metallic systems
in the Wigner-Dyson universality class of orthogonal symmetry $(\beta =
1)$.  The second term in the expression for $\tau$ is called the weak
localization correction.  It disappears in the limit $\epsilon\gg 1$,
where our field theory reduces to Efetov's nonlinear $\sigma$ model
with symmetry index $\beta = 2$.

Finally, in Section \ref{sec:strong} we tackle the problem of
exponential (or strong) localization, which emerges at lengths
$L/\xi_{\rm loc} > 1$.  The natural tool for calculating the
transmittance in this nonperturbative regime is the ``quantum
Hamiltonian'' of the one-dimensional functional integral.  By a
standard transfer matrix argument, the exponential decay of the
transmittance is determined by the lowest nonzero eigenvalue of the
quantum Hamiltonian.  Calculating it for an arbitrary energy
$\epsilon$ is a problem beyond the scope of this paper.  However, in
the two limits $\epsilon = 0$ and $\epsilon \gg 1$ the quantum
Hamiltonian is simply a Laplacian on a symmetric superspace and has a
large enough group symmetry to be tractable.  By computing the low
lying spectrum of the Laplacian we find that the localization length
for $\epsilon = 0$ is eight times smaller than for $\epsilon \gg 1$.

\section{Model and disorder averaged Green functions}
\label{sec:model}

Our starting point are the Bogoliubov-deGennes (BdG) equations for the
stationary quasiparticle excitations \cite{BdG} of a superconductor
in (BCS) mean field approximation:
      \begin{eqnarray*}
        H_0 u + \Delta v &=& E u \;, \\
        \Delta^* u - H_0^T v &=& E v \;.
      \end{eqnarray*}
The single-particle Hamiltonian $H_0$ is expressed by
      $$
      H_0 = (-i\nabla - eA)^2/2m + V(x) - \mu \;,
      $$ 
where $\mu$ is the chemical potential (Fermi level) and $A$ is the
magnetic vector potential.  The functions $u(x)$ and $v(x)$ describe
the particle and hole components of a quasiparticle, and are coupled
by the pairing potential $\Delta(x)$.  The potential energy $V(x)$ is
taken to be a Gaussian white noise random potential with zero mean and
variance

      $$
      \langle V(x)V(y)\rangle = v^2\delta(x-y) \equiv
      \frac{1}{2\pi\nu \tau} \; \delta(x-y) \;,
      $$
where $\nu$ is the density of states and $\tau$ is the elastic mean 
free time.  It is convenient to assemble $H_0$ and $\Delta$ into a
single operator $H$, here called the BdG ``Hamiltonian'' for short:

      $$
      H = \pmatrix{H_0 &\Delta\cr \Delta^* &-H_0^T \cr} \;,
      $$
which acts on the tensor product of ${\Bbb C}^2$ with the Hilbert
space of square-integrable functions on the coordinate space, $S$, of
the superconductor. The factor ${\Bbb C}^2$ will be referred to as the
particle-hole $(ph)$ space.

The BdG Hamiltonian has the following symmetry property:
      $$
      H = -{\cal C} H^T {\cal C}^{-1}
      \quad \mbox{with}\quad 
      {\cal C}= \pmatrix{0 &1\cr -1 &0\cr} \;,
      $$
where $T$ means transposition in both particle-hole and coordinate
space.  As a result of this symmetry, the eigenvalues of $H$ occur in 
pairs with opposite sign, and we have
      \begin{eqnarray*}
        G(z) &:=& (H-z)^{-1} = (-\CC H^T \CC^{-1}-z)^{-1} \\
        &=& -\CC \left( (H+z)^{-1} \right)^T \CC^{-1} = 
        -\CC G^T(-z) \CC^{-1} \;, 
      \end{eqnarray*}
so that the advanced and retarded Green functions are related by
      $$
      G_{\alpha\alpha'}^-(x,x';E) = - \sum_{\beta\beta'} 
      \CC_{\alpha\beta}\CC_{\alpha'\beta'}G_{\beta'\beta}^+(x',x;-E)\;.
      $$

To compute disorder averaged Green functions, we will use Efetov's
supersymmetry method \cite{efetov}, in its recent adaptation by
Altland {\it et al.} \cite{ats} to include the pairing correlations of
a superconductor.  In this method, disorder averages of Green
functions are obtained from a generating functional
      $$
      \CZ[j] = \Big\langle \int \CD \bar \psi \CD \psi \exp
      { \int_S d^3x \( i \bar \psi ( H - z ) \psi +
        \bar \psi j +\bar j \psi\)} \Big\rangle \;.
      $$
The functional integral is over supervectors $\psi$ and $\bar \psi$,
whose components are commuting and anticommuting (or Grassmann) fields
\cite{efetov}.  Thus, in addition to the (physical) particle-hole
space, an auxiliary ``boson-fermion'' $(bf)$ space is introduced.  To
average a product of $n$ Green functions, we further enlarge the field
space by forming the tensor product with ${\Bbb C}^n$ ({\it i.e.} we
take $n$ copies of the field space).  We do not distinguish between
``advanced'' and ``retarded'' fields, as we shall make use of the
aforementioned symmetry relating $G^-$ with $G^+$.

To take advantage of the particle-hole symmetry of the BdG
Hamiltonian, we find it convenient to tensor the field space with yet
another factor ${\Bbb C}^2$.  This extra ``spin'' degree of freedom,
referred to by the name of ``charge conjugation'' $(cc)$ space, is
introduced by rearranging the quadratic form of the generating
functional as follows:
      \begin{eqnarray*}
        2\bar \psi (H-z)\psi &=&
        \bar \psi (H-z)\psi + \psi^T
        (H^T -z) \bar \psi ^T\\ \nonumber
        &=&
        \bar \psi (H-z)\psi + \psi^T
        (-\CC^{-1}H \CC -z)\bar \psi ^T 
        \\ \nonumber
        &=&
        \pmatrix{\bar\psi & -\psi^T \CC^{-1}}
        \pmatrix{H-z &0\cr 0 &H+z\cr}
        \pmatrix{\psi\cr \CC\bar\psi^T\cr}\;.
      \end{eqnarray*}
The superscript $T$ here denotes the supertransposition operation. 
If we define
      $$
      \Psi=\frac{1}{\sqrt{2}}
      \pmatrix{\psi\cr \CC\bar \psi ^T\cr} \;,
      \qquad
      \bar\Psi = \frac{1}{\sqrt{2}}
      \pmatrix{\bar\psi & -\psi^T \CC^{-1}} \;,
      $$
we have 
      $$
      \bar\psi (H-z)\psi = \bar \Psi (H-\om)\Psi \;,
      $$
where $\om = \Sig_3 \otimes z$. We use the notation $\Sig_i$ for the
Pauli matrices acting in the charge conjugation space and $\sig_i$ for
the ones acting in the particle-hole space.  As is easily checked, the
two supervectors $\bar\Psi$, $\Psi$ obey the symmetry relations
      $$
      \Psi = \CC \gamma \,\bar \Psi^T \;,
      \qquad 
      \bar\Psi = - \Psi^T \CC^{-1} \gamma^{-1}\;,
      $$
where $\gamma$ is defined by
      $$
      \gamma = 1_{ph}\otimes\( E_{BB}\otimes \Sig_1 - 
      E_{FF}\otimes i\Sig_2\) \otimes 1_n \;
      $$
with $E_{BB}$ $(E_{FF})$ being the projector on the bosonic (fermionic)
space.  Note that $\ga$ and $\CC$ commute.  The trick of doubling
$\psi$ to $\Psi$ will help to make the symmetries of the nonlinear
$\sigma$ model manifest at all stages.

To summarize, the generating functional for averages of products of
Green functions can be written as
      \begin{equation}
        \label{generate}
        \CZ[J] = \Big\langle \int \CD \bar \Psi \CD \Psi \exp
        { \int_S d^3x \( i \bar \Psi ( H - \om)\Psi + 
          \bar \Psi J +\bar J \Psi \) } \Big\rangle
      \end{equation}
with sources $J$, $\bar J$ that obey the same symmetry relations as
$\Psi$, $\bar\Psi$.  The structure of the operators in this expression
is the following:
      \begin{eqnarray*}
        H &=& H_{ph}\otimes 1_{bf}\otimes 1_{cc} \otimes 1_n \;, \\
        \om &=& 1_{ph}\otimes 1_{bf}\otimes \Sig_3\otimes
        {\rm diag }(z_1,...,z_n) 
      \end{eqnarray*}
where $z_j = E_j - i\eta_j$ are complex numbers with negative
imaginary parts ($-\eta_j < 0$).  In using this tensor product
notation, we will usually omit trivial factors of unity.
  
After this digression about notation, we go about deriving the
effective field theory for the disordered superconductor. First of
all, decomposing $H$ into regular and stochastic parts as $H = {\cal
  H}_0 + V(x)\sigma_3$, we perform the average over disorder in the
generating functional $\CZ$:
      \begin{eqnarray*}
        \CZ[0] &=& \int \CD \bar \Psi \CD \Psi \int \CD V \exp{
          \int d^3x \;\( i \bar \Psi ( {\cal H}_0 - \om)\Psi
          + i \bar \Psi \sig_3 \Psi V(x)
          -\frac{V^2(x)}{2v^2}\)} \\
        &=& \int \CD \bar \Psi \CD \Psi  \exp{
          \int d^3x \;\( i \bar \Psi ( {\cal H}_0 - \om)\Psi-
          \frac{v^2}{2}( \bar \Psi \sig_3 \Psi)^2\)} \;.
      \end{eqnarray*}

The next step is to make a Hubbard-Stratonovich transformation, which
allows to extract the relevant ``slow modes'' of the problem.  We are
interested in keeping the diffusive modes, {\it i.e.} those modes that
arise from two-particle channels undergoing multiple scattering with
the exchange of momenta smaller than the inverse of the elastic
mean-free path, $\ell = v_F \tau$.  Isolating these modes is a
standard procedure \cite{efetov} which is conveniently performed in
Fourier space and has been included in Appendix \ref{sec:app_one} for
completeness.  It is shown there that the contribution from the slow
modes to $- v^2 (\bar\Psi \sig_3 \Psi)^2 / 2$ can be written as
      $$
      - v^2 \sum_{|q| < q_0} \str \zeta(-q) \zeta(q)
      $$
where $q_0 = 1/\ell$, and
      $$
      \zeta(q) = \sum_{k}  \Psi(k)\bar \Psi(-k+q) \sigma_3 \;.
      $$
One might think that the magnetic term in the Hamiltonian ${\cal H}_0$
suppresses the slow modes that couple to the vector potential $A$.  It
will turn out, however, that the magnetic field of a vortex is not
strong enough to cause complete suppression of these modes. Therefore,
we will keep all the slow modes appearing in the expression for
$\zeta(q)$.

To enable the treatment of the quartic term, one now introduces a 
Hubbard-Stratonovich like decoupling.  This procedure reduces the
quartic terms to bilinears, at the expense of introducing additional
fields $Q$:
      \begin{eqnarray*}
        &&\exp{\(-v^2\sum_q \str \zeta(q) 
          \zeta(-q)\)}=\\ \nonumber
        &&\int \CD Q \exp{\sum_q\(\frac{\pi \nu}{8 \tau}\str Q(q)Q(-q)
          -\frac{1}{2\tau}\str Q(q)\zeta(-q)\)}\;.
      \end{eqnarray*}
The symmetry relation
      \begin{eqnarray*}
        \str Q \Psi \bar\Psi \sig_3 
        &=& \str \sig_3 \bar\Psi^T \Psi^T Q^T
        = \str \sig_3 (\ga^{-1}\CC^{-1} \Psi)(- \bar\Psi \ga\;\CC) Q^T \\
        &=& -\str \CC \ga \;Q^T \ga^{-1} \sig_3 \CC^{-1} \Psi \bar \Psi 
        = \str \sig_1 \ga\; Q^T \ga^{-1} \sig_1 \Psi \bar\Psi \sig_3 \;,
      \end{eqnarray*}
is accounted for by subjecting the supermatrix $Q$ to the linear 
condition
      \begin{equation}
        Q = \sig_1 \ga \; Q^T \ga^{-1} \sig_1 \;.
        \label{qsymmetry}
      \end{equation}
After integrating out the quasiparticle fields $\Psi, \bar\Psi$ 
and switching back to the coordinate representation, we obtain
      \begin{eqnarray}
        \CZ[0] &=& \int \CD Q \; \exp - \CS[Q] \;, \nonumber \\
        \CS[Q] &=& - {\int d^3x \(\frac{\pi \nu}{8 \tau}\str Q^2 -
          \frac{1}{2}\str \ln ({\cal H}_0 - \om + i\sig_3 Q/2\tau) \)} \;.
        \label{action}
      \end{eqnarray}
The integration domain for $\Psi$ and $Q$ has to be chosen with care,
in order to ensure the convergence of all integrals.  A detailed
discussion how to choose integration manifolds correctly can be found
in \cite{mrz_rss}.

In summary, the problem of computing disorder averaged Green functions
has been reduced to considering the Euclidean field theory with
effective action $\CS[Q]$.  The sign difference compared to Efetov's
expression for the effective action comes from the different
definition of the supertrace.

\section{Saddle point approximations}
\label{sec:saddle}

The next step in deriving the nonlinear $\sigma$ model is to find the
saddle points of the effective action (\ref{action}), and to
incorporate the fluctuations around these.  Unfortunately, because the
gap function $\De$ varies in space, the full saddle point equation is
too complicated to solve directly.  However, in the dirty limit $\tau
\De \ll 1$, the minimum of the action functional can be found by a
two-step procedure \cite{ats}.  In that case the scales set by the
disorder and by the superconducting order parameter are well
separated, so that one can perform two minimizations in sequence.

To outline this section, the strategy adopted from \cite{ats} is as
follows.  At first, one neglects the gap function $\De$ and the
deviation $\omega$ of the energy from the Fermi level.  One varies the
resulting effective action and finds the corresponding saddle point
manifold.  Then, fluctuations inside this manifold are considered;
they couple to the gap function and to the energies $\om$.  The
resulting low-energy effective action is varied once again inside the
first (high-energy) saddle point manifold.  The partial differential
equation obeyed by the optimal configuration turns out to coincide
with the equation derived by Usadel \cite{usadel} for the Gorkov Green
function of a superconductor in the dirty limit.

\subsection{First saddle point}
\label{first}

Let us write the BdG Hamiltonian for a clean field-free superconductor
as
     $$
     {\cal H}_0\Big|_{A = 0} = h_0 \sig_3 + \tilde\Delta \;, \qquad
     \tilde\Delta = ({\rm Re}\De)\sig_1 - ({\rm Im}\De) \sig_2 \;,
     $$
where $h_0 = 
%{\nabla}
-\nabla^2/2m - \mu$.  In the absence of the gap and at the
Fermi energy, variation of the action functional $\CS[Q]$ gives the
following saddle point equation:
     \begin{equation}
       \label{speq}
       Q(x) =\frac{i}{\pi \nu} \langle x |(h_0+i\eta \sig_3 \Sig_3
       + iQ/2\tau)^{-1}|x \rangle \;,
     \end{equation}
where we kept the positive infinitesimal $\eta$, which is crucial for
distinguishing between the physical and unphysical solutions.  We
assume $\mu \gg 1/\tau$.  Then, since $\langle x | (h_0+i\eta)^{-1} |
x \rangle$ is understood as $-i\pi\nu$, equation (\ref{speq}) is
formally solved by a diagonal matrix $Q = {\rm diag} (q_1,q_2,...)$,
with $q_i = \pm 1$. To choose the signs correctly, we note that the
expression on the right-hand side of the saddle point equation relates
to the Green function for the disordered system in the self-consistent
Born approximation.  The disorder preserves the causal ({\it i.e.}
retarded versus advanced) character of the Green function, and
therefore the sign of $q_i$ must coincide with the sign of the
imaginary part of the energy.  This singles out the solution
      $$
      Q_0 =\sig_3 \Sig_3 \;.
      $$

\subsection{Fluctuations around the first saddle point}

In the limit $\tilde\Delta \to 0$, $A \to 0$ and $\om \to 0$, the
action (\ref{action}) is invariant under the transformation $Q(x)
\mapsto T Q(x) T^{-1}$, with $T$ any supermatrix that is constant in
space and consistent with the symmetry condition (\ref{qsymmetry}) for
$Q$.  Obviously, the saddle point equation must have the same
invariance, and its solution will therefore be degenerate.  Indeed,
the saddle points lie on a manifold that can be parametrized by $Q =
T Q_0 T^{-1}$, satisfying $Q^2 = 1$.  For $\mu\tau \gg 1$ the
low-energy excitations are given by slow transverse fluctuations
preserving the constraint $Q^2(x) = Q_0^2 = 1$.  To incorporate these,
we put $Q(x) := T(x) Q_0 T^{-1}(x)$ where $T(x)$ is determined only up
to ``gauge'' transformations $T(x)\mapsto T(x)h(x)$ that leave $Q_0$
fixed: $h(x) Q_0 h^{-1}(x)= Q_0$.  When taken modulo such gauge
transformations, the matrix $T(x)$ varies slowly with $x$.
Fluctuations that are not of this form are massive and will hence be
neglected.

We now insert $Q(x) = T(x) Q_0 T^{-1}(x)$ into the expression
(\ref{action}) for $\CS[Q]$ and expand with respect to $\tilde\De$ and
$\om$, and up to second order in gradients of $Q(x)$, neglecting
higher-order derivatives.  In this way the following {\it low-energy}
effective action governing the transverse fluctuations of $Q$ is
obtained:
      \begin{equation}
        \label{fseff}
        S[Q] = -\frac{\pi \nu}{8} \int d^3x\; 
        \str \( D(\nabla_A Q)^2 -4i (\tilde\De +\om)\sig_3 Q \) \;.
      \end{equation}
Here, $D = v_F^2\tau/3$ is the diffusion constant of the normal metal
and $\nabla_A = \nabla -ie A[\sig_3,\cdot]$.  The details of this
calculation can be found in Appendix \ref{sec:app_two}.

It will turn out to be of importance that, for $\om = 0$ (and any $A$
and $\tilde\Delta$), the action functional $S[Q]$ is invariant under
transformations
      \begin{equation}
        \label{invar}
        Q(x) \mapsto T Q(x) T^{-1}, \quad {\rm if} \quad 
        T = 1_{ph}\otimes t \;
      \end{equation}
and $t$ is constant in space and obeys $t = \gamma (t^{-1})^T
\gamma^{-1}$.  The latter condition means that $t$ runs through an
orthosymplectic Lie supergroup ${\rm OSp}(2n|2n)$.

\subsection{Second saddle point: Usadel equation}
\label{subsec:ssp}

We now want the optimal configuration of the superfield $Q$ for a
nonvanishing order parameter $\tilde\Delta$ and magnetic vector
potential $A$.  To find it, we require $S[Q]$ to be stationary with
respect to variations of $Q$ that preserve the constraint $Q^2 = 1$.
Such variations can be parametrized by
      $$
        \delta Q(x) = \epsilon [ X(x) , Q(x) ] \;,
      $$
where $X$ satisfies $X = - \sigma_1 \gamma X^T \gamma^{-1} \sigma_1$
so as to preserve the symmetry (\ref{qsymmetry}).  The stationarity 
condition $\delta S = 0$ then yields
      \begin{equation}
        \label{usadel}
        D \nabla_A \( Q \nabla_A Q \) + i[Q,(\tilde\De +\om)\sig_3] = 0\;.
      \end{equation}
With proper interpretation given to $Q$, this coincides with Usadel's
equation \cite{usadel} for the Gorkov Green function -- a fact first
understood in \cite{ats}.  By a slight abuse of terminology we shall 
refer to (\ref{usadel}) also as the Usadel equation.

Our plan now is as follows.  In the remainder of the current
subsection, we are going to solve the Usadel equation for a vortex
with cylindrical symmetry (at $\om=0$).  To do that, we will introduce
cylindrical coordinates $(r,\varphi,z)$ and look for a $z$-independent
solution $Q_0 (r, \varphi)$.  Because of the invariance (\ref{invar}),
this solution is degenerate and, once again, we are dealing with a
{\it manifold} of saddle points $T Q_0 (r, \varphi) T^{-1}$, generated
by the action of $T \in {\rm OSp}(2n|2n)$.  In the next section, we
will argue that the low-energy excitations of that manifold are given
by $Q(x) = T(z) Q_0( r,\varphi) T^{-1} (z)$, with $T(z) = 1_{ph}
\otimes t(z)$.  The fluctuations in the $r$ and $\varphi$ directions
are strongly suppressed by the coupling to the pairing gap of the
superconductor and by transverse quantization.  By inserting the last
form of $Q(x)$ into the effective action (\ref{fseff}) we will obtain
yet another effective action, this time for the matrix $T(z)$.

The first step is to solve the Usadel equation at $\omega = 0$.  This
was done before in \cite{usadelvortex}, and we redo it here in a way
tailored to our purposes.  Recall, first of all, the invariance of the
Usadel equation (at $\omega = 0$) under orthosymplectic
transformations $Q(x) \mapsto T Q(x) T^{-1}$, with $T = 1_{ph} \otimes
t$ and $t \in {\rm OSp}(2n|2n)$.  Now, far from the vortex, any
solution with finite action $S[Q]$ must become covariantly constant
$(\nabla_A Q \to 0)$ and adjust to the pairing gap $\tilde \Delta =
|\Delta| \sigma_1 {\rm e}^{-i\vphi\sig_3}$ of the superconductor in
the bulk $( [ Q , \tilde\Delta\sig_3 ] \to 0)$.  These conditions in
conjunction with $Q^2 = 1$ determine $Q$ at infinity to be $Q_\infty =
\sig_2 {\rm e}^{-i\varphi\sig_3}$, up to multiplication by a sign.
Note that $Q_\infty$ is stable under symmetry transformations:
$Q_\infty = (1_{ph}\otimes t) Q_\infty (1_{ph}\otimes t)^{-1}$.  On
the other hand, on approaching the vortex core, the unique value of
$Q_\infty$ is expected to {\it expand into a manifold} of solutions of
the Usadel equation, as a result of the orthosymplectic invariance.
To remove the resulting degeneracy, we hold the solution on the axis
of the vortex $(r = 0)$ fixed at the metallic saddle point $Q_0 =
\sig_3 \otimes \Sig_3$.  Given the two limiting forms at $r = 0$ and
$r = \infty$, it is natural to look for a solution of the form
      \begin{equation}
        \label{ansatz}
        Q(r,\varphi) = g(r) \sig_3 \otimes \Sig_3 +
        f(r) \sig_2 {\rm e}^{-i\vphi\sig_3}\otimes 1_{cc} \;,
      \end{equation}
which interpolates between $Q_0$ and $Q_\infty$. Note that this ansatz
is compatible with the symmetry condition (\ref{qsymmetry}).  To
fulfill the constraint $Q^2 = 1$, the functions $g(r)$ and $f(r)$ are
required to satisfy
      $$
      f^2(r) + g^2(r) = 1 \;.
      $$
Note also that $f(r)$ must vanish linearly at $r = 0$ in order for
$Q(r,\vphi)$ to be nonsingular there.

Next, we insert our ansatz into the Usadel equation, thereby obtaining 
a differential equation for $g(r)$ and $f(r)$.  As is shown in
Appendix \ref{sec:app_three}, this equation reads
      \begin{equation}
      {1 \over r} \partial_r (r f\partial_r g - r g\partial_r f) +
      \left( {1 \over r} - 2eA_\vphi \right)^2 fg 
      = {2 \over D} |\Delta| g \;.
      \label{diffeqn}
      \end{equation}
To complete the formulation of the problem, we need to specify the gap
function $|\Delta(r)|$ and the $\varphi$-component of the magnetic
vector potential, $A_\varphi(r)$.  In principle these quantities are
to be calculated self-consistently \cite{kramer} from the function
$f(r)$.  It is common knowledge, however, that their essential features
are as follows.  (i) The gap function vanishes on the symmetry axis of
the vortex, increases linearly with $r$ for small $r$, and saturates
at the bulk value of the superconductor over a scale given by the
coherence length $\xi$.  (ii) The magnetic field carried by the vortex
integrates to one fluxon $h/2e$ and is localized in a region of linear
size equal to the penetration depth $\lambda$.

For simplicity, we do not insist on an exact self-consistent solution
but adopt the following closed-form approximate expressions
\cite{blatter} (for an exact treatment, see Ref.~\cite{kramer}):
      $$
      2e A_\varphi(r) = {1 \over r} - \frac{K_1(r/\la)}{\la} \;, 
      \qquad |\De(r)| = \Delta_0 {r/\xi \over \sqrt{(r/\xi)^2+a^2} } \;,
      $$
which capture the features listed above.  The parameter $a$ is a
numerical constant whose value will be specified later on.  Notice
that the vector potential $A$ is everywhere nonsingular, as the Bessel
function $K_1 (r / \lambda)$ behaves as $\lambda / r$ near $r = 0$ to
cancel the pole $1 / r$.  Note also that, given the value of the gap
$\Delta_0$ far from the vortex, the coherence length $\xi$ no longer
is a parameter at our disposal.  Rather, $\xi$ has to be determined
self-consistently: because $\Delta$ is given by $f$ through the BCS
gap equation, the length scale $\xi$ fed into the differential
equation (\ref{diffeqn}) must agree with the characteristic scale of
the solution emerging from that equation.

To proceed, we solve the constraint $f^2 + g^2 = 1$ by setting $f =
\sin\theta$ and $g=\cos\theta$.  If we also introduce the dimensionless
variable $u = r / \xi$, the equation for $\theta(u)$ takes the form
      $$
      - {1 \over u} {d \over du} \left( u {d\theta \over du} \right)
      + \left( {K_1(u/\kappa) \over \kappa} \right)^2
      {\sin 2\theta \over 2} = {2 \xi^2 \Delta_0 \over D} \times
      {u \cos\theta \over \sqrt{u^2 + a^2} }
      $$
with $\kappa = \lambda/\xi$.  Let us now specialize to the case of an 
extreme type-II superconductor, where $\kappa \gg 1$ and the Bessel
function can be approximated by $K_1(u/\kappa)/\kappa \approx 1/u$.
In this limit, all explicit length scales are removed from the
equation -- and the self-consistency requirement of equal input and
output length scales is satisfied -- by setting $2 \xi^2 \Delta_0 / D
= {\rm const}$.  Without loss of generality we take the numerical
constant to be unity, so that 
      $$
      \xi = \sqrt{D / 2\Delta_0} \;.
      $$
(Note that a different choice of numerical constant would only amount 
to a redefinition of the parameter $a$.)  This result says that disorder
reduces the ``clean'' value of the coherence length $\xi_0 = v_F /
\pi\De_0$ to a ``dirty'' value $\xi \sim (\ell \xi_0)^{1/2}$.  The
Usadel equation then simplifies to
      \begin{equation}
        \label{trigonom}
        - \theta^{\prime\prime} - {\theta^\prime \over u} + 
        {\sin 2\theta \over 2u^2} = 
        {u\cos\theta \over \sqrt{u^2 + a^2}}\;.
      \end{equation}
As was stated before, the boundary conditions are $\cos\theta(\infty) 
= g(\infty) = 0$ and $\sin\theta(0) = f(0) = 0$.  Without loss of
generality we put $\theta(0) = 0$.

For large values of $u$, we can get an approximate analytical solution
as follows.  The boundary condition at infinity admits the possible
values $\theta(\infty) = (2m+1)\pi/2$ with $m$ an integer.  Writing
$\theta(u) = (2m+1)\pi/2 + h(u)$ and approximating $\sin h$ by $h$ in
the asymptotic regime, we get
      $$
      h^{\prime\prime}+ h^\prime / u = (-1)^m h \;.
      $$
The solutions of this second-order differential equation are
combinations of Bessel functions.  For even $m$, the solution that
goes to zero at infinity behaves asymptotically as
      $$
      h(u) \sim u^{-1/2} {\rm e}^{-u} \;,
      $$
while for odd $m$, the solutions have the asymptotic form
      $$
      h(u) \sim u^{-1/2} \cos (u + u_0) \;.
      $$
When inserted into the functional (\ref{fseff}), the latter solutions
can be shown to produce a radially integrated Lagrangian density
      $$
      \int_0^\infty g^2(r) r {d}r {d}\varphi \sim \int {d}r \;,
      $$
which diverges linearly.  On the other hand, the solutions for even
$m$, which decay exponentially, have a finite radial action.  Hence
the ``good'' boundary values at infinity are $\theta(\infty) = \pi/2 +
2\pi n$ with integer $n$.  This conclusion is confirmed by a second
argument: in Appendix \ref{sec:app_four} we show that only the
solutions with $\tan {1\over 2}\theta(\infty) > 0$ can be analytically
continued to the normal metal solution, which has to emerge at
energies large compared to the gap $\Delta_0$.

Since these solutions with $n \in {\Bbb Z}$ are local extrema (and, in
fact, minima) of the action, one might naively think that all of them
have to be taken into account, according to the principles of saddle
point evaluation of integrals.  However, the solutions with $n \not=
0$ do not appear anywhere in the large body of literature on the
Usadel equation and its applications.  What's the reason?  The answer
is that in the conventional kinetic theory the Usadel equation is
based on, one argues that the solutions with $n \not= 0$ are in
conflict with causality.  Indeed, $g(r)$ has an interpretation
\cite{usadel2} as the local density of states (LDoS) at radius $r$,
normalized to the density of states of the normal metal.  By the
properties of the cosine, the solution for $n = 0$ is expected to
yield a function $g = \cos\theta$ that is everywhere positive, but for
$n \not= 0$ regions of negative value inevitably appear.  In other
words, the $n \not= 0$ solutions fail to meet the physical requirement
of positivity of the density of states, or causality of the Green
function.

If this argument does not seem totally convincing -- on the grounds
that only the {\it total} (not the partial) LDoS is required to be
positive -- there exists also a functional integral reason why the $n
\not= 0$ solutions must be discarded.  A careful look reveals that for
$\theta(\infty) = \pi/2 + 2\pi n$ the field $Q$ winds $n$ times around
the singularity of the logarithm in equation (\ref{action}).
Consequently, the solutions with $n \not= 0$ are not smoothly
connected to any choice of integration manifold \cite{mrz_rss} for $Q$
(before saddle point approximation) that is permitted by the required
convergence of all integrals.  Put differently, the solutions with $n
\not= 0$ are not accessible by deformation of the integration manifold
without moving across singularities of the integrand.  This uniquely
singles out the solution with $n = 0$.  In summary, we adopt the two 
boundary conditions
      $$
        \th(0) = 0 \;, \quad {\rm and} \quad \th(\infty) = \pi/2 \;.
      $$

We have yet to fix the value of the parameter $a$ in equation
(\ref{trigonom}).  To do that, we follow Ref. \cite{usadelvortex}, 
and choose $a$ in such a way as to achieve exact self-consistency
far from the vortex.  Standard theory tells us that the order 
parameter $\Delta = \lambda \langle \psi_{\downarrow}
\psi_{\uparrow} \rangle$ (where $-\lambda$ is the BCS coupling
constant) is obtained by summing the thermal Green function
$\int_0^{\beta} \langle \psi_{\downarrow}(\tau) \psi_{\uparrow}(0)
\rangle {\rm e}^{i\omega\tau} {d}\tau$ over all Matsubara
frequencies $\omega_l = (2l+1) T/\pi$.  At temperature $T = 1/\beta =
0$ the sum condenses to an integral over the (imaginary) energy axis,
and by expressing the thermal Green function through the function
$f(u,E) = \sin\theta(u,E)$ one obtains
      \begin{equation}
        \Delta(u) / \Delta_0 = 
        \int_0^\infty \sin\theta(u,-i\omega) d\omega \Big/  
        \int_0^\infty \sin\theta(\infty,-i\omega) d\omega \;.
        \label{selfconsistency}
      \end{equation}
The function $\sin\theta(u,E)$ is determined by the Usadel equation at 
energy $E$ (see Appendix \ref{sec:app_three}):
      $$
      - \theta^{\prime\prime} - {\theta^\prime \over u} + 
      {\sin 2\theta \over 2u^2} = {u\cos\theta \over \sqrt{u^2 + a^2}}
      - i{E \over \Delta_0} \sin\theta \;.
      $$
The prescription of Ref.~\cite{usadelvortex} now is to expand the 
solution around the asymptotic limit
      $$
      \sin\theta(\infty,-i\omega) = 
      \Delta_0 / \sqrt{\Delta_0^2 + \omega^2} \;, \qquad
      \cos\theta(\infty,-i\omega) = 
      \omega / \sqrt{\Delta_0^2 + \omega^2} \;,
      $$
find the correction of order $1 / u^2$, insert the corrected solution
into equation (\ref{selfconsistency}), and equate the result for
$\Delta(u)$ with the ansatz $\Delta(u)/\Delta_0 = u / \sqrt{u^2 + a^2}
= 1 - a^2 / 2u^2 + ...$ made above.  This leads to a linear condition 
for $a$ giving $a^2 = \pi/2$.
\begin{figure}
        \hspace{2.0cm} \epsfxsize=9cm \epsfbox{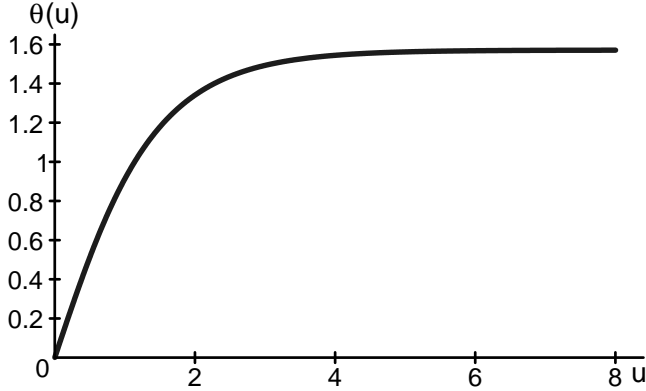}
        \caption{The solution $\theta(u)$ of the differential 
          equation (\ref{trigonom}) with the boundary conditions $\theta(0) 
          = 0$ and $\theta(\infty) = \pi/2$.  The value of the parameter 
          $a^2$ was chosen to be $\pi/2$.}
        \label{fig:vortex}
      \end{figure}

We have solved the differential equation (\ref{trigonom}) with $a^2 =
\pi/2$ numerically, using {\it Maple}.  Following again Ref.
\cite{usadelvortex}, we integrated the equation from $u = 0$ to $u =
\infty$, with the derivative $\theta^\prime(0)$ so adjusted as to hit
the desired value $\pi/2$ at infinity.  The solution is shown in
Figure 1.  In the next section we will see how this solution enters
into the expression for the coupling constant of the nonlinear
$\sigma$ model.

\section{The one-dimensional nonlinear $\sigma$ model}
\label{sec:onedim}

We are in the process of deriving a low-energy effective theory
describing diffusive quasiparticle motion in a disordered vortex.  The
final step, which we now embark on, begins by taking the solution of
the Usadel equation and perturbing it by fluctuations that vary slowly
in space.  Recalling $T(r,\varphi,z) = 1_{ph} \otimes t(r,\varphi,z)$,
we put
      $$
        Q = T Q_{\rm Usadel} T^{-1} = 
        g \sig_3 \otimes t \Sig_3 t^{-1} 
        + f \sig_2 {\rm e}^{- i\vphi\sig_3} \otimes 1 \;.
      $$
This expression is now inserted into the effective action
(\ref{fseff}).  Let us first evaluate the ``kinetic term'' $\str
(\nabla_A Q)^2$.  After some calculation we find that the magnetic
field disappears completely, the corresponding terms involving either
a trace over traceless matrices in the particle-hole space or a
supertrace over a supermatrix proportional to the identity.  We thus
obtain
      $$
      \str (\nabla_A Q)^2 = \str (\nabla Q)^2 \;.
      $$
The explicit form of the gradient is
      \begin{eqnarray*}
        \nabla Q &=& (\nabla g) \sig_3 t \Sig_3 t^{-1} 
        + g \sig_3 \nabla (t \Sig_3 t^{-1})
        + \nabla(f\sigma_2 {\rm e}^{- i\vphi\sig_3}) \;.
      \end{eqnarray*}
By the cyclic invariance of the supertrace and $\str 1 = 0$, the 
kinetic term reduces to
      \begin{eqnarray*}
        \str (\nabla Q)^2 &=& 
        g^2 \; \str_{ph\times bf\times cc\times n} 
        \( 1_{ph}\otimes \nabla (t \Sig_3 t^{-1}) \)^2 \\
        &=& 2 g^2 \; \str_{bf\times cc\times n} 
        \( \nabla (t \Sig_3 t^{-1}) \)^2 \;.
      \end{eqnarray*}
In the last expression the particle-hole space has disappeared, its
only remnant being the factor of 2 multiplying the restricted
supertrace.  By introducing a reduced $Q$-field,
      $$
      q = t \Sig_3 t^{-1} \;,
      $$
we can write more simply
      $$
      \str (\nabla Q)^2 = 2 g^2 \; \str (\nabla q)^2 \;.
      $$
The other term in $S[Q]$ works out to be 
      $$
      \str (\tilde\Delta + \omega)\sig_3 Q = 
      \str \tilde\Delta \sig_3 Q_{\rm Usadel} + 
      g \; \str \omega ( 1_{ph} \otimes t\Sigma_3 t^{-1}) 
      = 2 g \; \str \omega q \;. 
      $$
The effective action for the field $q$ then reads
      $$
      S[q] = -\frac{\pi\nu}{4} 
      \int d^3x \; \str\( g^2 D (\nabla q)^2 - 4 i g \omega q \) \;,
      $$

To identify the low-energy modes, we use cylindrical coordinates and
write the operator controlling the fluctuations around a constant
background $q$-field as
      $$
      - g^{-2} \nabla g^2 \nabla = - {\partial^2\over\partial z^2}
      - {1 \over r^2} {\partial^2 \over \partial\varphi^2} 
      - {1 \over r g^2(r)} {\partial \over \partial r} \left(
        r g^2(r) {\partial \over \partial r} \right) \;.
      $$
This operator is self-adjoint in the $L^2$-space with measure $g^2(r)
r{d}r {d}\varphi {d}z$.  The spectrum of the first term on
the right-hand side is continuous, while the second and third term
have a discrete spectrum.  [The latter is seen by making a similarity
transformation
      $$
      - {1 \over r g^2(r)} {\partial \over \partial r} \left(
        r g^2(r) {\partial \over \partial r} \right) \to
      - r^{-1/2} g^{-1}(r) {\partial \over \partial r}
      r g^2(r) {\partial \over \partial r} r^{-1/2} g^{-1}(r)      
      = - {\partial^2 \over \partial r^2} + V(r) \;,
      $$
and showing that the effective potential $V(r)$ is confining.]  Thus
the transverse, {\it i.e.} $r$ and $\varphi$, excitations of the
$q$-field are gapped (with the gap being of the order of the
transverse Thouless energy $D/\xi^2$), and the relevant low-energy
physics is captured by taking $q$ to be a function of $z$ only.  Doing
so we arrive at the following low-energy effective action for $q =
q(z)$:
      \begin{equation}
        \label{odsm}
        S[q] = -\frac{\pi^2 \nu \xi^2}{4} 
        \int {d}z\; \str\( {C_2 D}(\partial_z q)^2
        - 4 i C_1 \omega q \) \;,
      \end{equation}
where the dimensionless coefficients $C_1$ and $C_2$ are given by radial 
integrals,
      $$ 
      C_n = 2 \int_0^\infty \cos^n\th (u) \; u{d}u \qquad (n=1,2)\;,
      $$
which are of order unity.  To calculate their precise values one needs
the function $\cos\theta(u)$ which is determined by self-consistency.
If we take $\theta(u)$ to be the solution of equation (\ref{trigonom})
with $a^2 = \pi/2$, we get $C_1 = 3.16$ and $C_2 = 1.20$.  This
completes the derivation of the low-energy effective theory.  What we
have obtained is a one-dimensional field theory that belongs to the
general class of nonlinear $\sigma$ models.  The expression
(\ref{odsm}) for its action functional is the first important result
of the present paper.

To put the result into context, we elaborate on the definition of the
field space.  Recall that the field $q(z)$ is a $4n\times 4n$
supermatrix $q = t \Sig_3 t^{-1}$ with $t(z) \in {\rm OSp} (2n|2n)$.
Clearly, the expression for $q$ is invariant under right
multiplication of $t(z)$ by a matrix $h(z)$ that fixes $\Sig_3$ ($h
\Sig_3 h^{-1} = \Sig_3$).  By inspection, one easily sees that the
group of such matrices is isomorphic to ${\rm GL}(n|n)$.  Thus, the
field space of our nonlinear $\sigma$ model is a quotient ${\rm
  OSp}(2n|2n) / {\rm GL}(n|n)$.  Let us emphasize that the present
theory is {\it distinct} from the nonlinear $\sigma$ models that were
introduced by Efetov \cite{efetov} in his work on disordered metallic
wires.  In fact, what we are dealing with here is one of the new
``normal-superconducting'' universality classes (namely class $C$)
identified in Ref. \cite{az_ns}.  To demonstrate the distinction
clearly, let us take a close-up look at the field space.

Since $\gamma = ( E_{BB}\otimes \Sig_1 - E_{FF}\otimes i\Sig_2 )
\otimes 1_n$ gives $\Sig_3 = - \gamma (\Sig_3)^T \gamma^{-1}$, the Lie
group condition $t = \gamma (t^{-1})^T \gamma^{-1}$ translates into $q
= - \gamma q^T \gamma^{-1}$.  A useful way of representing $q = t
\Sig_3 t^{-1}$ is by
      $$
      q = \pmatrix{ (1+Z\tilde Z)(1-Z\tilde Z)^{-1}
                    &-2Z(1 - \tilde Z Z)^{-1} \cr
                    2\tilde Z (1-Z\tilde Z)^{-1} 
                    &-(1+\tilde Z Z)(1-\tilde Z Z)^{-1}\cr} \;,
      $$
where $Z$ and $\tilde Z$ are complex $2n \times 2n$ supermatrices.
The condition $q = - \gamma q^T \gamma^{-1}$ is equivalent to
      $$
      Z = \pmatrix{Z^{BB} &Z^{BF}\cr Z^{FB} &Z^{FF}\cr} = 
      \pmatrix{-(Z^{BB})^T &-(Z^{FB})^T\cr -(Z^{BF})^T &(Z^{FF})^T\cr}\;,
      $$
      $$
      \tilde Z = \pmatrix{\tilde Z^{BB} &\tilde Z^{BF}\cr 
        \tilde Z^{FB} &\tilde Z^{FF}\cr} = 
      \pmatrix{-(\tilde Z^{BB})^T &(\tilde Z^{FB})^T\cr 
        (\tilde Z^{BF})^T &(\tilde Z^{FF})^T\cr} \;.
      $$
We see in particular that $Z^{FF}, \tilde Z^{FF}$ are symmetric
matrices, while $Z^{BB}, \tilde Z^{BB}$ are skew.  Moreover, stability
of the functional integral with action (\ref{odsm}) requires
      $$
      \tilde Z^{FF} = - (Z^{FF})^\dagger \;, \quad
      \tilde Z^{BB} = + (Z^{BB})^\dagger \;, \quad
      1 - (Z^{BB})^\dagger Z^{BB} > 0 \;.
      $$
When complemented by the ${\rm OSp}(2n|2n)$ action induced on $Z,
\tilde Z$ by $q \mapsto t q t^{-1}$, the above properties mean that
$Z^{BB}$ and $Z^{FF}$ parametrize certain symmetric domains that are
well-known in mathematical physics.  Technically speaking, the
$BB$-sector is isomorphic to the noncompact symmetric space ${\rm
  SO}^*(2n)/{\rm U}(n)$ (called type $D{\rm III}$ in Cartan's
notation), while the $FF$-sector is isomorphic to the compact
symmetric space ${\rm Sp}(2n)/{\rm U}(n)$ (called type $C{\rm I}$).
In the terminology of \cite{mrz_rss}, these facts are succinctly
summarized by saying that the field space of our nonlinear $\sigma$
model is a Riemannian symmetric superspace of type $D{\rm III}|C{\rm
  I}$.  In contrast, Efetov's models are defined over the Riemannian
symmetric superspaces of type $A{\rm III}|A{\rm III}$, $BD{\rm
  I}|C{\rm II}$, and $C{\rm II}|BD{\rm I}$.

Before turning to the analysis of the one-dimensional field theory,
let us briefly discuss its zero-dimensional limit.  Such a limit is
obtained if the vortex is taken to be isolated, with Neumann boundary
conditions $(\partial_z q)(0) = (\partial_z q)(L) = 0$ at the two end
points $z = 0$ and $z = L$. (The boundary conditions on the
microscopic quasiparticle wave functions at an insulating boundary are
of course Dirichlet, but the reduction to the effective field theory
turns them into Neumann.)  At small energies $\omega$, the dominant
contribution to the functional integral comes from the ``zero mode''
$q(z) = q_0$ (independent of $z$).  Spatially varying field
configurations have a Lagrangian density that scales with the length
of the vortex as $L^{-2}$.  By comparing terms in the action
functional, one sees that they are negligible if
      $$
      \omega \ll D/L^2 \;.
      $$
This condition, which delineates the zero-dimensional (or ergodic)
regime, is familiar from diffusive metals:  ergodic behavior sets
in at times much longer than the diffusion time $L^2 / D$ or,
equivalently, at energies much smaller than the Thouless energy
$E_{\rm Th} = D/L^2$. 

In the zero-dimensional limit, the functional integral over $q(z)$
reduces to a definite integral over the zero mode $q_0$:
      \begin{eqnarray*}
        {\cal Z}[J] &=& \int Dq_0 \; \exp \left( -i\pi \str \omega q_0 
          / 2 \delta + {\rm sources} \right) \;, \\
        1/\delta &=& 2\pi  C_1 \nu L \xi^2 \;.
      \end{eqnarray*}
To obtain any specific observable, one keeps track of the coupling to
the source terms introduced in the starting expression
(\ref{generate}).  For the special case of the $n$-level correlation
function, the corresponding definite integral was computed for all $n$
in \cite{mrz_circular}.  The result is
      \begin{equation}
        \langle \nu(E) \rangle = {1\over\delta} - 
        {\sin 2\pi E/\delta \over 2\pi E} \;
        \label{density}
      \end{equation}
for the mean density of states as a function of the energy difference
$E$ from the Fermi level, and
      $$
      Y_2(\varepsilon_1,\varepsilon_2) = \left( 
        {\sin \pi(\varepsilon_1 -\varepsilon_2) \over \pi
          (\varepsilon_1 - \varepsilon_2)}
        - {\sin \pi(\varepsilon_1 + \varepsilon_2) \over \pi
          (\varepsilon_1 + \varepsilon_2)}
      \right)^2 \qquad (\varepsilon_i = E_i / \delta) \;
      $$
for the two-level cluster function $Y_2$.  These results agree exactly
with the expressions predicted in \cite{az_ns} on the basis of a
random matrix hypothesis for systems of symmetry class $C$.

How does this compare with the results expected on the basis of the
traditional approach due to Eilenberger and Usadel?  The Usadel
equation by itself, {\it i.e.} without the nonperturbative extension
worked out here, simply yields $1/\delta$ for the mean density of
states at energies comparable to the level spacing.  While this answer
is correct asymptotically at $E \gg \delta$, it ignores the
oscillations seen in (\ref{density}) and, in particular, it misses the
fact that the density of states vanishes at $E = 0$.  (Note that the
Usadel equation predicts secular variations of the density of states
over a characteristic scale given by the transverse Thouless energy
$D/\xi^2$.  For our purposes, this is a ``high-energy'' scale which
we shall ignore.)

Why does random matrix theory predict the fine structure of the
density of states correctly, whereas the quasiclassical Usadel
equation captures only the asymptotics?  Perhaps the answer is not
entirely obvious and the following remark helpful.  What is well
understood is that random matrix theory provides a valid description
of disordered systems in the {\it ergodic} regime, where the physics
is governed by the zero modes or {\it constant} modes of some field
theory. [Recall that the ergodic regime is the region of validity of
(\ref{density}).]  One might object, however, that a vortex is a
spatially localized object.  The solution of the Usadel equation for a
disordered vortex is far from being constant but has a characteristic
spatial dependence. One might therefore wonder why a zero mode or
random matrix argument has any chance of being correct in the present
context.

Our analysis has already given the answer to this question: although
the solution of the Usadel equation does vary in space, {\it the
  symmetry group operating on it is independent of space}!  As we have
seen, the random matrix answer is recovered by integrating over the
orbit swept out by the (constant) action of the symmetry group.  A key
feature is the mechanism by which the theory avoids the superficial
conflict between constancy of the zero mode and locality of the vortex
solution: the orbit generated by the group action {\it contracts to a
  single point} $Q_\infty = T Q_\infty T^{-1}$ (thereby making the
symmetry ineffective) as we move away from the vortex.

To conclude this section, let us mention that similar statements apply
also to an SNS junction with phase difference $\phi_1 - \phi_2 = \pi$
(symmetry class $C$I).  In that case, too, the random matrix
predictions of \cite{az_ns} are expected to hold as they stand.

\section{Transmission along the vortex}
\label{sec:trans}

Charge transport through a diffusive metallic wire at zero temperature
undergoes a crossover from Ohmic behavior to strong localization as
the length of the wire is increased.  A similar crossover is expected
for the transport properties of the normal quasiparticles of a
disordered vortex.  In the case of metallic wires, the precise form of
the crossover functions for the electrical conductance was computed
from the solution of Efetov's nonlinear $\sigma$ models in one
dimension in \cite{mrz_fourier,mmz}.  (Note that an error made in
these computations for $\beta = 4$ was corrected in \cite{bf}.)  We
will now embark on a similar project for the vortex, by analysing the
one-dimensional field theory found above.  Note that for the purpose
of testing for localization of normal quasiparticles in a vortex,
charge transport is not an experimentally viable option, as the system
is short cut by dissipationless charge flow through the
superconductor.  However, we can imagine probing the transport
properties via a measurement of the heat flow along the vortex.
Whether such a measurement is experimentally feasible will be
discussed in detail in a separate publication.

In this section we will show that the normal quasiparticles in a
disordered vortex are subject to a phenomenon called ``weak
localization'', which is well known from disordered metals with
time-reversal symmetry and conserved electron spin.  Although in
normal metals weak localization is destroyed by the application of a
magnetic field, in a disordered vortex there exist additional modes of
quantum interference that arise from Andreev scattering of the
quasiparticles and are not suppressed by the presence of a magnetic
flux.  The zero-dimensional precursor of weak localization in systems
of symmetry class $C$ was first noticed in a transfer-matrix
calculation by Brouwer and Beenakker \cite{bb} and was qualitatively
explained in \cite{az_andreev}.

As a quantitative measure of quasiparticle transport and localization
in a disordered vortex, we study the following two-particle Green
function:
      $$
      \tau_{kk'}(x,x';E) = 
      \sum_{\alpha\alpha'} \left\langle v_{k\alpha}^{(x)} 
        G^+_{\alpha\alpha'} (x,x';E) v_{k'\alpha'}^{(x')} 
        G^-_{\alpha'\alpha}(x',x;E) \right\rangle \;,
      $$
where $v$ is the left-right symmetric velocity operator
      $$
      v_{k\alpha} = {1 \over 2im} \left( (\sigma_3)_{\alpha\alpha}
        \left({\buildrel \rightarrow \over {\partial\over\partial x^k}} - 
         {\buildrel \leftarrow \over {\partial\over\partial x^k}}\right) - 
         2ie A_k\right) \;.
      $$
The superscripts $\pm$ distinguish between retarded and advanced
Green functions, and $\alpha , \alpha'$ are the particle/hole indices.
The angular brackets indicate an average over disorder.  

If $\psi = \{\psi_\alpha\}_{\alpha = {p,h}}$ is a solution of the
BdG equations (with some energy $E$), the bilinear $j_k(x) =
\sum_\alpha (\bar\psi_\alpha v_{k\alpha} \psi_\alpha)(x)$ expresses
the $k^{\rm th}$ component of the probability current carried by a
quasiparticle with wavefunction $\psi$.  This current is conserved
($\sum_k \partial_k j_k = 0$), as an immediate consequence of the BdG
equations.  On the same grounds, the tensor $\tau_{kk'}$ satisfies the
current conservation laws
      $$
      \sum_k {\partial \over \partial x^k} \; \tau_{kk'}(x,x';E) 
      = 0 = \sum_{k'} {\partial \over \partial {x'}^{k'}} \; 
      \tau_{kk'}(x,x';E) \;.
      $$
This suggests that $\tau_{kk'}(x,x';E)$ can be interpreted as a
correlation function of the conserved probability flux carried by
quasiparticles with energy $E$.  

The nonlocal tensor $\tau_{kk'} (x,x'; E)$ carries more information
than we are actually interested in here.  To eliminate some of the
details, let $P_z$ and $P_{z'}$ be two planes that intersect the axis
of the vortex at right angles, and consider the integrated
longitudinal correlation function
      $$
      \tau(E) := \int_{P_z} d^2x \int_{P_{z'}} d^2x'
      \; \tau_{33}(x,x';E) \;.
      $$
By current conservation, this is independent of $z$ and $z'$.  In
fact, $\tau(E)$ is a measure of the ``transmittance'' of the vortex
for quasiparticles of energy $E$, and plays a similar role as the
electrical conductance in a metallic wire.

To set up the functional integral for $\tau(E)$, recall the relation
      $$
        G^-_{\al'\al} (x',x;E) = - \sum_{\beta\beta'}
        \CC_{\alpha\beta} \CC_{\alpha'\beta'} G^+_{\beta\beta'}(x,x';-E) \;,
      $$
which follows from the particle-hole symmetry of the BdG Hamiltonian.
Introducing two replicas and choosing to represent the Green functions
in the bosonic sector, we write
      \begin{eqnarray*}
        \tau_{kk'}(x,x';E) & = & \sum_{\alpha\alpha'\beta\beta'}
        \CC_{\alpha\beta} \CC_{\alpha'\beta'}\left\langle \left\langle 
         v_{k\alpha}^{(x)} \psi^{B1}_\alpha(x) \bar\psi^{B1}_{\alpha'}(x') 
         \; v_{k'\alpha'}^{(x')} \psi^{B2}_\beta(x) \bar\psi^{B2}_{\beta'}(x')
        \right\rangle \right\rangle \\
        & = & -~4 \sum_{\alpha\alpha'} \left\langle \left\langle 
        \bar\Psi^{B22}_\alpha(x) v_{k\alpha} \Psi^{B11}_\alpha(x) \times
        \bar\Psi^{B11}_{\alpha'}(x') v_{k'\alpha'} \Psi^{B22}_{\alpha'}(x') 
        \right\rangle \right\rangle \;,
      \end{eqnarray*}
where the second line introduces the charge conjugation space,
by making use of the relations established in Section \ref{sec:model}.
The multiple superscript comprises the boson-fermion, replica and
charge conjugation indices, in this order.  The double angles mean an
average over disorder and over the configurations of the fields
$\Psi(x)$.  The symmetry breaking term in the action functional is
given by
      $$
      \omega = 1_{ph}\otimes 1_{bf}\otimes \Sig_3 \otimes E \sig_3 \;.
      $$

Now, after Hubbard-Stratonovich transformation, integration over the
quasiparticle fields and saddle point approximation, we obtain
      $$
      \tau_{kk'}(x,x';E) = - \pi^2 D^2 \nu^2 \left\langle 
        {\cal J}_k^{B11,B22}(x) {\cal J}_{k'}^{B22,B11}(x')\right\rangle \;,
      $$
where 
      $$
      {\cal J} = \sum_\alpha (Q \nabla_A Q)_{\alpha\alpha} \;.
      $$
For a consistency check on this calculation, note that the Usadel
equation yields
      $$
      \sum_k {\partial \over \partial x^k} \; {\cal J}_k^{B11,B22}(x) = 0 = 
      \sum_k {\partial \over \partial x^k} \; {\cal J}_k^{B22,B11}(x) 
      \;,
      $$
in agreement with the current conservation law satisfied by the tensor
$\tau_{kk'}(x,x';E)$.

The last step is to use the parametrization of Section \ref{sec:onedim}
for the matrix $Q$:
      $$
      Q = g(r) \sig_3 \otimes t(z) \Sig_3 t^{-1}(z) 
        + f(r)\sig_2 {\rm e}^{- i\vphi\sig_3} \;,
      $$
and to integrate over the planes $P_z$ and $P_{z'}$.  We then obtain
      $$
      \tau(E) = c_0 \left\langle j^{B11,B22}(z) j^{B22,B11}(z')
      \right\rangle
      $$
where $j = q \partial_z q$ and
      $$
      c_0 = - 4 \pi^2 D^2 \nu^2 \( \int g^2 (r) r {d} r 
      {d} \varphi\)^2 = - (2\pi^2 C_2 \nu D \xi^2)^2 \;.
      $$

For a diffusive vortex of short length, $L$, we expect the
transmittance $\tau(E)$ to obey Ohm's law ($\tau \sim 1/L$).  Our
goal now is to verify this law and compute the weak localization
correction, {\it i.e.} the next term in an expansion in the ratio of
system size over the localization length.  To that end, we write the
action functional (\ref{odsm}) in rescaled form:
      $$
        S[q]= -\frac{\xi_{\rm loc}}{16L}\int_0^1 {d}s \; 
        \str \( (\partial_s q)^2 - 2i \ep \lambda q\)\;,
      $$
$s$ being the dimensionless coordinate along the vortex.  The ratio
$L/\xi_{\rm loc}$, where
      $$
      \xi_{\rm loc} = 4\pi^2 C_2 \nu D \xi^2
      $$
will later be identified as a localization length, plays the role of the
coupling constant of the one-dimensional field theory.  The parameter
$\ep$,
      $$
      \ep = \frac{2\,C_1 E L^2}{C_2 D} \;,
      $$
represents the quasiparticle energy measured in units of the Thouless 
energy $E_{\rm Th} = D/L^2$, and $\lambda$ is the matrix
      $$
      \lambda = 1_{bf} \otimes \sigma_3 \otimes \Sigma_3 \;.
      $$
In keeping with what was said in the introduction, the functional
integral is supplemented by open boundary conditions at the two ends
of the vortex:
      $$
      q(0) = \Sigma_3 = q(1) \;.
      $$
It is well understood that these are the correct boundary conditions
to use when the system is in good contact with infinite particle
reservoirs.  (Roughly speaking, the deviation of $q(z)$ from
$\Sigma_3$ measures how often a typical Feynman path of the
quasiparticle visits the location $z$.  A diffusive path that arrives
at one of the end points $s = 0$ or $s = 1$, most likely exits the
vortex and disappears into the large phase space of the reservoir,
with vanishing probability of return.  This draining effect by the
coupling to the reservoirs causes the fluctuations of $q$ away from
$\Sigma_3$ to be suppressed at an open boundary.)  On moving the two
planes $P_z$ and $P_{z'}$ to the ends of the vortex ($z = 0$, $z' = L$
or $s = 0$, $s' = 1$), the expression for the transmittance $\tau$
becomes
      $$
      \tau(E) = - \left(\frac{\xi_{\rm loc}}{2}\right)^2
      \left\langle j^{B11,B22}(0) j^{B22,B11}(1) \right\rangle \;.
      $$

\subsection{Parametrization of $q$}
\label{sec:q}

Recall that the supermatrix $q$ satisfies the orthosymplectic
condition
      $$
        q = - \ga q^T \ga^{-1} \;, \quad {\rm with} \quad
        \ga = \left( E_{BB}\otimes\Sigma_1 - 
          E_{FF}\otimes i\Sigma_2 \right) \otimes 1_2 \;.
      $$
The kinetic term $\str (\partial q)^2$ in the action functional
$S[q]$ is invariant under global transformations $q(z) \mapsto t q(z)
t^{-1}$ with $t \in {\rm OSp}(4|4)$.  This symmetry is broken by the
potential term $\str \lambda q$ at finite energy $\epsilon \not= 0$.
We will see that increasing the symmetry breaking coupling $\epsilon$
causes crossover to Efetov's nonlinear $\sigma$ model with unitary
symmetry (or $\beta = 2$), thereby eliminating the weak localization
correction in the limit $\epsilon\to\infty$.

The effect of symmetry breaking is best captured by parametrizing $q$
as
      $$
        q = t_0 t_1 \Sigma_3 (t_0 t_1)^{-1} \;,
      $$
where the matrices $t_0$, $t_1$ decompose into blocks in charge 
conjugation space as follows:
      $$
        t_0 = \pmatrix{1 &Z_0 \cr \tilde Z_0 & 1\cr} \;,
        \qquad
        t_1 = \pmatrix{1 &Z_1 \cr \tilde Z_1 & 1\cr} \;.
      $$
The $4 \times 4$ blocks $Z_0, \tilde Z_0, Z_1, \tilde Z_1$ are 
chosen to have the following substructure in replica space:
      \begin{eqnarray*}
        Z_0 & = & \pmatrix{0 &z_0^+\cr z_0^- &0\cr} \;,
        \qquad
        Z_1 = \pmatrix{z_\wedge &0\cr 0 &z_\vee\cr} \;, \\
        \tilde Z_0 & = & \pmatrix{0 &\tilde z_0^+\cr \tilde z_0^- &0\cr} \;,
        \qquad
        \tilde Z_1 = \pmatrix{\tilde z_\wedge &0\cr 0 &\tilde z_\vee\cr} \;.
      \end{eqnarray*}
Here all quantities $z_\wedge, z_\vee, z_0^+, z_0^-$ and their
partners with tilde are $2\times 2$ supermatrices with symmetries that
are determined by $q = - \gamma q^T \gamma^{-1}$.  The rationale
behind this choice is that $t_0$ commutes with $\lambda$ ($t_0^{-1}
\lambda t_0 = \lambda$), so that the symmetry breaking term reduces to
      $$
      \str \lambda q = \str \lambda t_0 t_1 \Sigma_3 (t_0 t_1)^{-1}
      = \str \lambda t_1^{\vphantom{-1}} \Sigma_3^{\vphantom{-1}}
      t_1^{-1} \;,
      $$
which simplifies the calculation appreciably.  A similar
parametrization was used by Altland, Iida and Efetov \cite{ida} in
their study of the spectral statistics of small metallic particles
subject to a symmetry breaking magnetic field.  The boundary
conditions at $s = 0,1$ translate into
      $$
      Z_0(0) = \tilde Z_0(0) = Z_0(1) = \tilde Z_0(1) = 0,
      $$
with identical relations in force for $Z_1, \tilde Z_1$.  Using these,
we can linearize the current $j = q\partial_z q$ in $Z_0,Z_1$ at the
boundaries, and the expression for the transmittance becomes
      $$
      \tau = - (\xi_{\rm loc}/L)^2 
      \left\langle \partial_s z_0^{+BB}(0) \;
        \partial_s \tilde z_0^{-BB}(1) \right\rangle \;.
      $$

\subsection{Expansion of the kinetic and potential terms}

We will now use standard field theoretic perturbation theory to
compute the leading terms of the Taylor expansion in $L/\xi_{\rm loc}$
for $\tau$.  The first step is to expand the action functional $S[q]$
up to fourth order in the fields.  Setting $q = t_0 t_1 \Sigma_3 (t_0
t_1)^{-1}$ we obtain for the kinetic part of the action
      $$
      \str(\partial_s q)^2 = \str \left[ 
        t_1^{-1} \partial_s^{\vphantom{-1}} t_1^{\vphantom{-1}} +
        t_1^{-1} (t_0^{-1}\partial_s^{\vphantom{-1}} t_0^{\vphantom{-1}})
        t_1^{\vphantom{-1}} \; , \; \Sigma_3 \right]^2 \;.
      $$
By inserting the explicit expressions for $t_0$ and $t_1$ in terms of
$Z_0, \tilde Z_0$ {\it etc.} and using the fact that $Z_0$ and $Z_1$
are orthogonal to each other with respect to the supertrace, we get for
$\str (\partial_s q)^2$ up to second order
      $$
        -8\str\left( \partial\tilde Z_0\partial Z_0 +
          \partial\tilde Z_1\partial Z_1 \right) \;,
      $$
and a fourth order contribution given by
      \begin{eqnarray}
        \label{fourthorder}
        & - & 8\str \Big( 
          \partial{Z}_0\partial\tilde{Z}_0
          ({Z}_1 \tilde{Z}_1 + {Z}_0 \tilde{Z}_0) 
          + \partial\tilde{Z}_0\partial Z_0 
          (\tilde{Z}_1 Z_1 + \tilde Z_0 Z_0)
          \Big) \nonumber \\
        & + & 8\str \left( Z_0 \partial\tilde{Z}_0 
          (Z_1\partial\tilde Z_1 - \partial Z_1 \tilde Z_1) +
          \tilde Z_0\partial Z_0 (\tilde Z_1 \partial Z_1 -
            \partial \tilde Z_1 Z_1) \right) \\
        & + & 8\str \left( (\partial Z_0\tilde Z_1)^2
          + (\partial \tilde{Z}_0 Z_1)^2 -
          \partial \tilde Z_1\partial Z_1 \tilde{Z}_1 Z_1 -
          \partial Z_1 \partial \tilde Z_1 Z_1 \tilde Z_1 \right) 
        \nonumber \;.
      \end{eqnarray}
For simplicity we have set $\partial \equiv \partial_s$.  For the
potential term we have the expansion
      $$
      \str \lambda q = 4 \str
        (\tilde z_\wedge z_\wedge - \tilde z_\vee z_\vee)
      + 4 \left(\str (\tilde{z}_\wedge z_\wedge)^2 - 
        \str(\tilde{z}_\vee z_\vee)^2 \right) + ...\;,
      $$
which depends only on the matrix entries of $Z_1, \tilde Z_1$.

\subsection{Propagator and diagrams}

The quadratic part of the field theory action defines the bare
propagator. We write
      $$
        S[q] = - \frac{\xi_{\rm loc}}{16L} \int_0^1 {d}s \; 
        \str \left( (\partial q)^2 - 2i\epsilon \lambda q \right)
        = S_2 + S_4 + \ldots
      $$
where
      $$
        S_2 = \frac{\xi_{\rm loc}}{2L} \int_0^1 {d}s \; 
        \left( \str(\partial\tilde Z \partial Z) + i\epsilon 
          \str(\tilde z_\wedge z_\wedge - \tilde z_\vee z_\vee)\right) \;,
      $$
and $Z$ in this expression means $Z_0+Z_1$.  From the quadratic part
of the action it is clear that the modes of type $Z_0$ are massless,
while the modes of type $z_\vee$ and $z_\wedge$ have an imaginary mass
term given by $\ep$.  In the limit $\ep\to\infty$, the latter modes
will therefore be completely suppressed.

We have yet to discuss the symmetries of the matrices $Z$ and $\tilde
Z$.  It is easy to see by linearization that the symmetry $q = -
\gamma q^T \gamma^{-1}$ is equivalent to the condition
      $$
      X = - \gamma X^T \gamma^{-1} \quad {\rm on} \quad
      X = \pmatrix{0 &Z\cr \tilde Z &0\cr} \;.
      $$
On decomposing $\gamma$ into blocks in charge conjugation space,
      $$
      \gamma = (E_{BB}\otimes\Sigma_1 - E_{FF}\otimes i\Sigma_2)
      \otimes 1_2 = \pmatrix{0 & (\sigma_3)_{bf}\otimes 1_2   \cr 
      1_2\otimes 1_2    &0\cr} \;,
      $$
we see that the $4\times 4$ supermatrices $Z = Z_0 + Z_1$ and $\tilde
Z = \tilde Z_0 + \tilde Z_1$ must obey
      $$
      Z = -\sigma  Z^T \;, \quad \tilde Z = - \tilde Z^T \sigma  \;,
      $$
where $\sigma = (\sigma_3)_{bf} \otimes 1_2$ is the superparity matrix,
with value $+1$ on bosons and $-1$ on fermions.  Recalling the 
definition of the supertranspose $T$ and denoting the superparity of 
$a$ by $|a|$, we equivalently have
      $$
      Z^{ab} = - (-1)^{|a||b|} Z^{ba} \;,
      \quad
      \tilde Z^{ab} = - (-1)^{|a| + |b| + |a||b|} \tilde Z^{ba} \;.
      $$
In this notation, the kinetic term reads $\str \partial\tilde Z
\partial Z = \sum_{ab} (-1)^{|a|} \partial\tilde Z^{ab}\partial
Z^{ba}$.

\begin{figure}
        \hspace{2.0cm} \epsfxsize=10cm \epsfbox{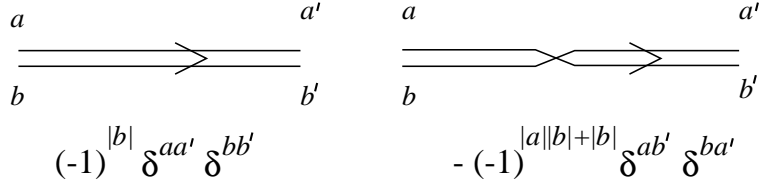}
        \caption{Graphs for the propagator 
         $\langle Z^{ab} \tilde Z^{b'a'} \rangle$}
        \label{fig:propag}
      \end{figure}

The basic Gaussian integral governing the perturbation expansion is
given by
      \begin{eqnarray*}
        \left\langle Z^{ab}(s) \tilde Z^{b'a'}(s') \right\rangle_0 
        &:=& \int {\cal D}Z{\cal D}\tilde Z\; Z^{ab}(s)\tilde Z^{b'a'}(s') 
        \; \exp \left( \frac{\xi_{\rm loc}}{2L} \int_0^1 {d}s \sum_{cd}
          (-1)^{|c|} \tilde Z_{cd} \partial^2 Z_{dc} \right) 
        \\ \nonumber 
        &=& (-1)^{|b|} \left( \delta^{aa'}\delta^{bb'} - 
          (-1)^{|a||b|}\delta^{ab'}\delta^{ba'} \right) 
        (L/\xi_{\rm loc}) (- \partial^2)^{-1}(s,s') \;.
      \end{eqnarray*}
This formula applies to the massless modes $Z_0$.  For the modes $Z_1$
({\it i.e.} $z_\wedge$ and $z_\vee$) one needs to replace $-\partial^2$
by $-\partial^2+i\epsilon$ and $-\partial^2-i\epsilon$ respectively.
The index structure on the right-hand side is determined by the super
skewsymmetry of the matrices $Z$ and $\tilde Z$.  It is seen that
there are two different ways of propagating the matrix indices, which
we represent graphically as shown in Figure \ref{fig:propag}. 

The space dependence of the propagator is given by
      $$
      {\rm D_0}(s,s') := (-\partial^2)^{-1}(s,s') =
      \left\{ \begin{array}{cc}
          s(1-s') & \; {\rm for} \ \ s \leq s' \;, \cr
          s'(1-s) & \; {\rm for} \ \ s \geq s' \;,
        \end{array} \right.
      $$
for the massless modes, and by
      \begin{eqnarray*}
      {\rm D}_\wedge(s,s') &:=& (-\partial^2 + i\epsilon)^{-1}(s,s') \\
      &=&
      \frac{1}{\sqrt{i\ep} \sinh \sqrt{i\epsilon}}
      \left\{ \begin{array}{cc}
          \sinh \sqrt{i\epsilon} s \; \sinh \sqrt{i\epsilon}(1-s')
          & \; {\rm for} \; s \leq s' \;, \cr
          \sinh \sqrt{i\epsilon} s' \; \sinh\sqrt{i\epsilon} (1-s)
          & \; {\rm for} \; s \geq s' \;,
        \end{array} \right.
      \end{eqnarray*}
for those with mass squared equal to $i\epsilon$.  Note that these Green
functions satisfy Dirichlet boundary conditions at the ends $s = 0$
and $s = 1$, as required.  The Green function ${\rm D}_\vee(s,s') =
(-\partial^2 - i\epsilon)^{-1}(s,s')$ is obtained from ${\rm
  D}_\wedge(s,s')$ by the substitution $\epsilon \to -\epsilon$.

We are now ready to calculate the transmittance of the vortex up
to one-loop order.  What we want is the quantity
      $$
      \tau = - (\xi_{\rm loc}/L)^2 \left\langle \partial Z_0^{B1,B2}(0)
      \partial\tilde Z_0^{B2,B1}(1) \times (1-S_4) \right\rangle_0\;,
      $$
where $S_4$ is the part of the action functional that is quartic in
$Z$ and $\tilde Z$, and the vacuum expectation value is taken with
respect to the Gaussian action $S_2$.  $S_4$ contains terms from the
kinetic and the potential part in $S$.

Let us first compute the leading contribution (tree level).  Using the 
expression for the massless propagator we obtain
      \begin{eqnarray*}
      \tau_0 &=& -(\xi_{\rm loc}/L)^2 \left\langle \partial Z_0^{B1,B2}(0)
      \partial\tilde Z_0^{B2,B1}(1) \right\rangle_0 \\
      &=& - (\xi_{\rm loc}/L) \partial_{s} \partial_{s'} 
      {\rm D}_0(s,s')\Big|_{s=0,s'=1} = \xi_{\rm loc}/L \;.
      \end{eqnarray*}
This is the equivalent of Ohm's law for the electrical conductance.
Note that $\partial_s \partial_{s'} {\rm D}_0(s,s')$ is in fact
independent of $s$ and $s'$, as is expected from conservation of the
quasiparticle probability flux.  The Ohmic limit $\tau_0 = \xi_{\rm
  loc} / L$ gives support to the interpretation of $\xi_{\rm loc}$ as a
localization length.  In a diffusive wire one would have $\xi_{\rm
  loc} \sim N\ell$ with $N$ the number of channels open at the Fermi
energy or, equivalently, $\xi_{\rm loc} \sim \nu {\cal A} D$ with
${\cal A}$ the cross sectional area of the wire.  In the present case
we saw that
      $$
      \xi_{\rm loc} \sim \nu \xi^2 D \;.
      $$
This means that the role of the cross sectional area is here taken
by $\xi^2 \sim D / \De_0 \sim \xi_0 \ell$.

We are now going to compute the weak localization correction, which is
of order $(L/\xi_{\rm loc})^0$.  First, we notice that the ``external
legs'' $\partial Z_0^{B1,B2}(0)$ and $\partial\tilde Z_0^{B2,B1}(1)$
can only couple to terms containing at least one $Z_0$ and one $\tilde
Z_0$.  Therefore the fourth order terms coming from the potential, as
well as the terms on the third line in (\ref{fourthorder}), can be
omitted from $S_4$ for present purposes.  In Appendix
\ref{sec:app_five} we show that the contributions originating from the
terms on the second line of (\ref{fourthorder}) cancel each other, so
we need only consider the terms coming from the first line:
      $$
      \tau_1 = \frac{\xi_{\rm loc}^3}{2L^3} \left\langle
      \partial Z_0^{B1,B2}(0) \partial\tilde Z_0^{B2,B1}(1)
      \int_0^1 {d}s \; \str (
      Z \tilde Z \partial Z_0 \partial\tilde Z_0 +
      \tilde Z Z \partial\tilde Z_0 \partial Z_0 
      ) \right\rangle \;,
      $$
where we again used the notation $Z = Z_0 + Z_1$ and the fact that
$Z_0$ and $Z_1$ are orthogonal with respect to the supertrace.

\begin{figure}
        \hspace{1.8cm} \epsfxsize=10cm \epsfbox{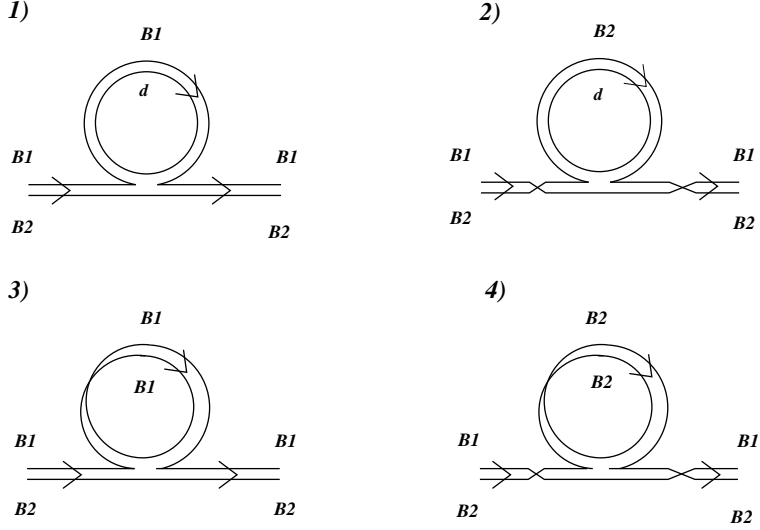}
        \caption{One-loop diagrams potentially contributing to the 
          transmittance.}
        \label{fig:diagram}
      \end{figure}

Next, we draw the one-loop diagrams that potentially contribute to the
transmittance.  The ones associated with the term $\str Z\tilde
Z\partial Z_0 \partial\tilde Z_0$ are displayed in Figure
\ref{fig:diagram}.  Those associated with $\str \tilde Z Z \partial
\tilde Z_0 \partial Z_0$ are similar, except that the loop is placed
in the lower part of the graph.  The first two diagrams, which contain
a freely running index, vanish by the usual mechanism of
supersymmetry: the fermion loops come with a minus sign relative to
the boson loops.

Let us now compute the contribution from Diagram $3$. In the course of 
this computation it will become clear that the diagrams with massless 
loops (type $Z_0$) are zero.  First we consider the following contraction 
of the fields:
      $$
      \sum_{abcd} (-1)^{|a|} \; \left\langle \underbrace{
          \partial Z_0^{B1,B2}(0) \underbrace{
            \partial\tilde Z_0^{B2,B1}(1) \; 
            \partial Z_0^{ab}} \; \partial\tilde Z_0^{bc}}
        \underbrace{\; Z^{cd} \; \tilde Z^{da}} \Big|_s
        \right\rangle \;, 
      $$
which yields the expression
      $$
      \sum_{d} 
      \left\langle 
        \partial Z_0^{B1,B2}(0) \partial\tilde Z_0^{B2,B1}(s)
      \right\rangle_0
      \left\langle 
        \partial Z_0^{B1,B2}(s) \partial\tilde Z_0^{B2,B1}(1)
      \right\rangle_0
      \left\langle 
        Z^{B1,d}(s) \tilde Z^{d,B1}(s)
      \right\rangle_0 \;.
      $$
The propagator associated with the loop, $\langle Z^{B1,d}(s) \tilde
Z^{d,B1}(s) \rangle_0$, consists of two terms, one containing a factor
$(-1)^{|d|} \delta^{ B1,B1} \delta^{dd}$ (Diagram 1) and another one
with $-\delta^{B1,d} \delta^{d,B1}$ (Diagram 3).  The first term
vanishes by supersymmetric cancellation, $\sum_d (-1)^{|d|} = 0$, as
stated above.  (Note in particular that all contributions from
$Z_0$-loops cancel by this mechanism.)  The second gives
      $$
      -\partial_{s_1}\Big|_{s_1 = 0}\partial_s {\rm D}_0(s_1,s) 
      \; {\rm D}_\wedge(s,s) 
      \; \partial_s \partial_{s_2}\Big|_{s_2 = 1}{\rm D}_0(s,s_2) \;.
      $$
Diagram 4 gives a very similar contribution, the only difference being
that $D_\vee(s,s)$ replaces $D_\wedge(s,s)$.  It should also be clear
now that if the external legs are contracted with the fields $Z$ and
$\tilde Z$, the loop is of $Z_0$ type and yields a vanishing
contribution.

The same result is obtained when the diagrams arising from $\str
\tilde\partial Z_0 \partial Z_0 \tilde Z Z$ are considered.  Although
there might in principle be extra contributions from the expansion of
the functional integration measure, we show in Appendix
\ref{sec:app_six} that this is not the case.

In summary, we obtain the following result for the weak localization
correction to the transmittance:
      \begin{eqnarray*}
        \tau_1 &=& - \int_0^1 {d}s \; \partial_{s_1} \partial_s
        {\rm D}_0(s_1,s)\Big|_{s_1=0} \( {\rm D}_\wedge(s,s) + 
        {\rm D}_\vee(s,s) \) \;
        \partial_{s} \partial_{s_2} {\rm D}_0(s,s_2)\Big|_{s_2=1} \\ 
        &=& - \int_0^1 {d}s \( \frac{\sinh \sqrt{i\ep}s\; \sinh 
          \sqrt{i\ep}(1-s)}{\sqrt{i\ep} \; \sinh \sqrt{i\ep}}+
        \frac{\sinh \sqrt{-i\ep}s \; \sinh \sqrt{-i\ep}(1-s)}
        {\sqrt{-i\ep}\;\sinh\sqrt{-i\ep}} \) \\ 
        &=& - {1\over 2} \(\frac{\coth \sqrt{i\ep}}{\sqrt{i\ep}}+
        \frac{\coth \sqrt{-i\ep}}{\sqrt{-i\ep}}\)\;.
      \end{eqnarray*}
Using the expansion of the hyperbolic cotangent around $x = 0$,
      $$
      \coth x = \frac{1}{x} + \frac{x}{3} + \ldots \;,
      $$
we find that the weak localization correction for quasiparticles of
zero energy is $\tau_1(\ep=0) = -1/3$.  In the limit $\ep \to \infty$
the correction vanishes as $1/\sqrt{\ep}$.  The crossover between the
two regimes takes place at $\ep \sim 1$, corresponding to energies of
the order of the Thouless energy $E_{\rm Th} = D/L^2$.

To finish this section, let us briefly recall the physical picture
\cite{az_ns} explaining the origin of the weak localization correction
computed above.  Weak localization in normal-/superconducting systems
with a magnetic field arises from modes of interference that are
present already in the {\it single-particle} Green function.  The
relevant diffusive processes are described by single-particle Feynman
paths containing a closed loop which is traversed {\it twice}.  (This
is the physical interpretation of the third and fourth diagram in
Figure \ref{fig:diagram}.)  To make such a process phase coherent in
the presence of a magnetic field, the charge state of the
quasiparticle during the second traversal must be opposite to its
state during the first traversal (the charge is reversed by Andreev
reflection).  In that case, the total magnetic phase accumulated in
the loop vanishes, and the process survives disorder averaging.  The
net effect of such processes is to suppress the density of states and
reduce the transmittance of the vortex (weak localization).

The preceding discussion assumes the energy of the quasiparticle to be
zero or negligibly small.  At finite energy $E$, the phase coherence
between the two charge states in the closed loop is reduced by an
additional dynamical phase, which is proportional to $E$ multiplied by
the time typically spent in the loop.  For diffusive dynamics, this
time is estimated to be of the order of $L^2/D$ in an open system of
size $L$.  Therefore, phase decoherence due to $E > 0$ becomes
effective at energies of the order of the Thouless energy $D/L^2$.

In the limit $E \gg D/L^2$, states of opposite charge can no longer
interfere, so the problem reduces to considering BdG particles only,
or BdG holes only.  The surviving modes of interference are the same
as those of a metallic wire subject to a magnetic field.  It is well
known that the latter system has a vanishing weak localization
correction, and one must go to two-loop order to see the onset of
localization.  The field theory describing such systems is Efetov's
nonlinear $\sigma$ model with unitary symmetry ($\beta = 2$).  What
one expects, then, is that the field space of $q(z)$ becomes
isomorphic to Efetov's model space $(\beta = 2)$ when the modes $Z_1$,
which are killed by the limit $E\to\infty$, are set to zero.
Such an isomorphism in fact exists, and will be constructed in the
next section.

\section{Strong localization}
\label{sec:strong}

To introduce the subject of this section, we briefly recall how the
conductance of a disordered quasi-one-dimensional normal metal varies
with the length $L$ of the system \cite{reviews}.  Provided that $L$
is not too small, the disorder averaged conductance $g(L)$ depends
only on a single scaling variable:
      $$
      g(L) = F_\beta (L/\xi_{\rm loc}) \;,
      $$
where $\xi_{\rm loc}$ is a characteristic scale for the onset of
localization.  The scaling function $F_\beta$ is specific to each
Wigner-Dyson universality class (with symmetry index $\beta = 1$, $2$,
or $4$).  Let the normalization conventions for $\xi_{\rm loc}$ be
such that $F_\beta$ has the Ohmic (or small $x$) limit $F_\beta(x) =
1/x + ...$ in all cases.  Then, to next order in the small-$x$
expansion,
      $$
      F_\beta(x) = {1 \over x } - c_\beta + {\cal O}(x) \;,
      $$
where the constant $c_\beta$, the weak localization correction, is
$c_\beta = 2/3$, $0$, $-1/3$ for $\beta = 1$, $2$, $4$.  Looking back
at the result of the previous section, we see that weak localization
for the low-energy quasiparticles of a disordered vortex is
qualitatively the same as for metallic wires with symmetry index
$\beta = 1$.

In the opposite limit of large $x$ the scaling functions fall off 
exponentially:
      $$
      F_\beta(x) \stackrel{x\to\infty}{\longrightarrow}
      A_\beta \exp ( - x / \beta ) ,
      $$
which is an expression of the phenomenon of strong localization that
occurs in quasi-one-dimensional systems.  Similar behavior is expected
for the transmittance $\tau$ of the vortex.  The purpose of the
present section is to confirm this expectation and work out the
precise value of the exponent governing the exponential decay law.

Recall the formula expressing $\tau$ as a functional integral over
fields $Z$:
      $$
      \tau = - (\xi_{\rm loc}/L)^2 
      \left\langle \partial_s Z_0^{B1,B2}(0) \;
        \partial_s \tilde Z_0^{B2,B1}(1) \right\rangle \;.
      $$
For the purpose of extracting the large-$L$ asymptotics, it is best to
abandon the functional integral formulation and pass to an equivalent
quantum mechanical description.  To do that we interpret the
dimensionless coordinate $s$ as an imaginary time.  Since the
Lagrangian of our system is 
      $$
      (\xi_{\rm loc} / 16 L) \; \str \left( - (\partial_s q)^2 
        + 2i\epsilon \lambda q \right) \;,
      $$ 
the evolution with ``time'' ({\it i.e.} length) will be governed by
the Hamiltonian
      $$
      {\cal H} = - {L \over 2\xi_{\rm loc}} \; \triangle + 
      {i\epsilon \xi_{\rm loc} \over 8L} \; \str \lambda q \;,
      $$
where $\triangle$ is the Laplace-Beltrami operator of the field space
$G_2 /H_2 = {\rm OSp}(4|4)/{\rm GL}(2|2)$ with metric tensor $- \str
({\rm d}q)^2 / 8$.  The open boundary conditions on the fields mean
that we want a quantum transition amplitude from $Z = 0$ at $s = 0$ to
$Z = 0$ at $s = 1$.  Moreover, by the rule that $\dot x$ for a
Euclidean path integral with Lagrangian density $m \dot x^2 / 2 +
V(x)$ translates into the differential operator $m^{-1} d/dx$ in the
quantum theory, the expression for $\tau$ becomes
      $$
      \tau = \left\langle Z=0 \left| D_0^{B1,B2} {\rm e}^{-{\cal H}}  
          \tilde D_0^{B2,B1} \right| Z=0 \right\rangle \;,
      $$
where $D_0^{B1,B2} = \partial / \partial\tilde Z_0^{B2,B1}$ and
$\tilde D_0^{B2,B1} = \partial / \partial Z_0^{B1,B2}$. 

Supersymmetry guarantees the ground state of ${\cal H}$ to be exactly
at zero.  By inserting a complete set of eigenstates of ${\cal H}$, we
see that the exponential decay of $\tau$ for large $L$ is determined
by the smallest eigenvalue of ${\cal H}$ such that the corresponding
eigenfunction $\psi$ has nonzero matrix element $\langle 0 |
D_0^{B1,B2} | \psi \rangle$.  Unfortunately, computing the spectrum
(or even the lowest nonzero eigenvalue) of ${\cal H}$ for all
$\epsilon$ is a rather difficult task.  The only cases we can do
easily are $\epsilon = 0$ and $\epsilon = \infty$.  These are the
cases we will restrict ourselves to here.

Consider the case $\epsilon = 0$ first.  The Hamiltonian we are facing
is minus the Laplacian on $G_2 / H_2$.  Because the Laplacian commutes
with the action of the symmetry group $G_2$, its spectrum is in
principle computable by the powerful machinery of Lie group theory.
There exists, however, a trick we can use to further simplify the
problem.  Although we are forced to introduce {\it two} replicas to
average a product of one retarded and one advanced Green function when
the quasiparticle energy $E$ is nonzero, in the limit $E = 0 =
\epsilon$ we can make do with just a {\it single} replica.  The idea
is to use again the BdG particle-hole symmetry to express $G^-(E)$ by
$G^+(-E)$, and then to represent one Green function in the bosonic
sector and the other one in the fermionic sector.  This can be done
without breaking the supersymmetry of the formalism if and only if $E
= 0$.  Developing the rest of the theory in the same manner as before,
we arrive at a reduced description in terms of a ``small'' $q$ field
taking values in $G_1 / H_1 = {\rm OSp}(2|2) / {\rm GL}(1|1)$.  (The
external derivatives $D_0^{B1,B2}$ and $\tilde D_0^{B2,B1}$ in the
transition amplitude formula for $\tau$ are replaced by odd
derivatives $D^{BF}$ and $\tilde D^{FB}$.)

To obtain the lowest nonzero eigenvalue of minus the Laplacian on $G_1
/ H_1$ with a minimal amount of mathematical effort, we adopt the
pedestrian approach of choosing a suitable coordinate system and
expressing the Laplacian as a differential operator in those specific
coordinates.  A particularly useful coordinate system is provided by
the following polar decomposition:
      \begin{eqnarray*}
        q &=& h a \Sigma_3 a^{-1} h^{-1} \;, \quad
        a = \exp \pmatrix{0 &Y\cr Y &0\cr} \;, \quad
        Y = \pmatrix{0 &0\cr 0 &i\theta\cr} \;, \\
        h &=& \exp \pmatrix{h_1^T &0\cr 0 &h_1^{-1}\cr} \;, \quad
        h_1 = \exp \pmatrix{0 &\zeta\cr \bar\zeta &0\cr}
        \exp \pmatrix{0 &0\cr 0 &i\varphi\cr} \;.
      \end{eqnarray*}
Here $\theta, \varphi$ are real commuting variables, and $\zeta,
\bar\zeta$ are complex anticommuting.  We refer to $\theta$ as the
``radial'' coordinate.  The metric tensor in this coordinate system
takes the form
      $$
      - \str ({\rm d}q)^2 / 8 = {\rm d}\theta^2 + \sin^2 2\theta \;
      \left( {\rm d}\varphi + \bar\zeta{\rm d}\zeta/2i + 
        \zeta{\rm d}\bar\zeta/2i \right)^2 + \sin^2 \theta \; 
      {\rm d}\zeta {\rm d}\bar\zeta \;.
      $$
Now, there exists a standard formula for the Laplacian on a Riemannian 
(super)manifold:
      $$
      \triangle = \sum_{ij} (-1)^{|i|} 
      g^{-1/2}(\xi) \partial_{\xi^i} g^{ij}(\xi)
      g^{1/2}(\xi) \partial_{\xi^j} \;,
      $$
if the metric tensor is $\sum {\rm d}\xi^i g_{ij}(\xi) {\rm d}\xi^j$,
and $g(\xi) = {\rm SDet}(g_{ij}(\xi))$.  Given this formula one can
write down a lengthy expression for the Laplacian on $G_1 / H_1$.
However, there is yet another simplification and we do not need to go
through the trouble of writing the full expression.  The point is that
we can construct the desired eigenfunction of $\triangle$ by repeatedly
acting with symmetry generators on a corresponding {\it radial}
eigenfunction $f(\theta)$.  It is therefore sufficient to study the
spectrum of the radial Laplacian, which from the above expression for
the metric is
      $$
      \triangle^\# = 
      \left( {\sin 2\theta \over \sin^2 \theta} \right)^{-1}
      {d \over d\theta}
      \left( {\sin 2\theta \over \sin^2 \theta} \right)
      {d \over d\theta} \;.
      $$
The low lying spectrum of this operator can be constructed very simply
by trial and error.  Making the substitution $x = \cos 2\theta$ we
have
      $$
      - \triangle^\# / 4 = - (1 - x) {d \over dx} (1 + x) {d \over dx} \;,
      $$
and it is easy to verify that this has the eigenfunctions
      $$
      f_0 = 1 \;, \quad
      f_1 = 1-x \;, \quad
      f_2 = (1-x)(1+3x) \;,
      $$
with eigenvalues $\lambda_0 = 0$, $\lambda_1 = 1$, and
$\lambda_2 = 4$.  The pattern we see suggests that the complete
spectrum of $- \triangle^\# / 4$ might be $n^2$ with $n$ an integer.
(This guess is in fact confirmed by more informed Lie algebraic
considerations.)  In any case, it is clear that the lowest nonzero
eigenvalue is $\lambda_1 = 1$.  By applying to $f_1 = 1-x$ an odd
generator of the symmetry group $G_1$, one can explicitly construct an
odd eigenfunction $\psi$ of $-\triangle/4$ with eigenvalue +1 and matrix
element $\langle 0 | D^{BF} | \psi \rangle \not= 0$.  Since the
Hamiltonian is ${\cal H} = - L \triangle / 2 \xi_{\rm loc}$, we
predict the decay of the transmittance to be
      $$
      \tau(E = 0) \sim \exp \left( - 2 L/\xi_{\rm loc} \right)
      $$
for quasiparticles with zero energy.

The other limit $E\to\infty$ can be done more quickly, by drawing on
known results.  As we saw, this limit eliminates the modes $Z_1$, and
what remains are the modes $Z_0$, which are recalled to have the
matrix representation
      $$
      Z_0 = \pmatrix{0 &z_0^+\cr z_0^- &0\cr} \;, \quad
      \tilde Z_0 = \pmatrix{0 &\tilde z_0^+\cr \tilde z_0^- &0\cr} \;.
      $$
By the off-diagonal structure of these matrices, the field space
decomposes into two separate pieces, which are isomorphic copies of
each other by the orthosymplectic relations $Z_0 = -\sigma Z_0^T $ and
$\tilde Z_0 = -\tilde Z_0^T \sigma$.  We can write the Lagrangian in
terms of one of these copies, say $z_0^+$ and its partner $\tilde
z_0^-$, and account for the other one by multiplying by 2.  The
kinetic term $-(\xi_{\rm loc}/16L) \; \str (\partial q)^2 \big|_{Z_1 =
  0}$ then takes the form
      $$
      (\xi_{\rm loc} / L) \; \str
      (1 - \tilde z_0^- z_0^+)^{-1} \partial\tilde z_0^-
      (1 - z_0^+ \tilde z_0^-)^{-1} \partial z_0^+ \;.
      $$
Recall that $z_0^+$ and $\tilde z_0^-$ are complex $2 \times 2$
supermatrices unconstrained by any symmetry relations that do not
involve complex conjugation.  This means that the object at hand is
the Lagrangian of Efetov's nonlinear $\sigma$ model for metallic wires
of the Wigner-Dyson symmetry class $\beta = 2$, in the specific
parametrization given by
      $$
      Q = \pmatrix{1 &z_0^+\cr \tilde z_0^- &1\cr}
      \pmatrix{1 &0\cr 0 &-1\cr}
      \pmatrix{1 &z_0^+\cr \tilde z_0^- &1\cr}^{-1} \;,
      $$
and with a renormalized value of the localization length.  This model
has been much studied in the literature \cite{larkin,mrz_fourier,mmz}.
Taking the result from there, we find
      $$
      \tau(E = \infty) \sim \exp \left( - L/4\xi_{\rm loc} \right) \;.
      $$
Thus the localization length at $\epsilon = 0$ is predicted to be
shorter than at $\epsilon \gg 1$ by a large factor of 8.  This
strong energy dependence already announced itself in the weak
localization correction, and now we see that it becomes even more
dramatic in the region of strong localization.

\section{Concluding remarks}

In this paper we have established a field theoretic description of the
diffusive motion of normal quasiparticles along the core of a vortex
in a dirty, extreme type-II, s-wave superconductor.  The technique
used was an extension, due to Ref.~\cite{ats}, of the supersymmetric
method developed by Efetov and others for the purpose of studying
localization and mesoscopic phenomena in normal metals.  A
comprehensive account of this extension and its application to
normal-/superconducting mesoscopic systems will appear shortly
\cite{ast}.

In a simplified picture, one may view a vortex core as a normal
conducting wire surrounded by superconducting walls.  The
quasiparticles in such a system undergo Andreev scattering and partake
in interference processes that are not present in a disordered normal
metal.  This gives rise to different spectral correlations and
transport properties.  For a short, isolated vortex the correlations
of the quasiparticle spectrum (in the universal low-energy limit) have
been verified to agree with those of a random matrix of symmetry class
$C$.  For an open system and quasiparticle energies below $E_{\rm Th}
= D/L^2$, the ``transmittance'' defined in Section \ref{sec:trans}
exhibits a weak localization correction.  Strong localization sets in
for vortices of length $L > \xi_{\rm loc}$.  The localization length
$\xi_{\rm loc}$ for a quasiparticle with energy above the Thouless
energy $E_{\rm Th}$ is eight times larger than the localization length
of a low-energy quasiparticle.

These results were obtained by reducing the effective field theory for
the quasiparticles to a one-dimensional nonlinear $\sigma$ model.  In
the process, we had to solve a saddle point equation, which turned out
to be the Usadel equation.  The saddle point approximation is valid if
$\mu\gg 1/\tau$ and $1/\tau \gg \Delta_0$, the latter being known as
the dirty limit.  The reduction to one dimension (the degree of
freedom along the vortex) can be done for quasiparticle energies
smaller than the transverse Thouless energy $D/\xi^2$.

It is a well known fact that nonmagnetic, isotropic disorder has no
significant effect on s-wave superconductivity (Anderson's theorem).
This ceases to be true when the order parameter is not spatially
homogeneous.  Then, a self-consistent field treatment is required.
The self-consistency problem for the disordered vortex has been
studied both in the framework of Landau-Ginzburg theory \cite{BdG},
and by using the Usadel equation \cite{kramer}.  Both approaches show
that disorder with an elastic mean-free path $\ell$ changes the
coherence length from its clean value $\xi_0$ to a dirty value $\xi
\sim (\ell \xi_0)^{1/2}$.  We have seen that this scaling of the dirty
coherence length comes about quite naturally, by using no more than
dimensional analysis and some minimal assumptions about the shape of
the condensate.

We did not implement an exact self-consistent treatment here. Instead,
we took the order parameter profile $\Delta(r)/\Delta_0$ from a simple
one-parameter family of functions and chose the free parameter so as
to achieve self-consistency in the region of approach to the
asymptotics far from the vortex.  We believe that an exact treatment
will lead only to minor corrections in the numerical constants $C_1$
and $C_2$ that enter the coupling constants of the nonlinear $\sigma$
model.  In any event, the qualitative picture that emerges from our
results, namely, a crossover between the low-energy regime with weak
localization present in the ``transmittance'' along the vortex, and
the high-energy regime without weak localization, is a robust feature
that will not be modified by a more quantitative treatment.

For typical values of a superconductor with a normal metallic
diffusion coefficient of the order of $20{\rm cm}^2/{\rm sec}$, and a
critical temperature around $10{\rm K}$, a rough estimate of the
localization length gives a range $1-10{\rm mm}$.  This corresponds to
an elastic mean-free path of the order of $5{\rm nm}$ and a coherence
length of about $100{\rm nm}$.  For a sample of $1-10\mu{\rm m}$
thickness, the Thouless energy of the vortex will be in the ${\rm mK}$
range.  Because of the large localization length, the weak
localization correction is a relatively small effect.  Still, the
predicted weak localization correction to the transmittance along the
vortex should be visible as a low-temperature anomaly in the heat
conductivity of an ensemble of vortices between two electron
reservoirs.  We shall discuss this effect in detail in a forthcoming
communication.

{\bf Note added.} After completion of this manuscript, we learned of
related work by Skvortsov, Kravtsov, and Feigel'man \cite{skf}.

{\bf Acknowledgment.} We are much indebted to Alexander Altland for
valuable discussions and input at many stages of the present work, and
for locating Ref. \cite{usadelvortex}.  The initial spark for this
research came from Gerd Bergmann, who a number of years ago asked one
of us (M.R.Z.) how the disordered vortex relates to the conventional
classification scheme for disordered electron systems.

\newpage
\appendix
\section{Decoupling of the slow modes}
\label{sec:app_one}

Consider the quartic term 
      $$
      {v^2 \over 2} \int d^3x 
      \left( \bar\Psi(x) \sigma_3 \Psi(x) \right)^2 \;,
      $$
which appears on averaging over the random potential $V(x)$.  We want
to isolate in it the collective modes with small momentum, $|q| < q_0$,
which are to be decoupled by a Hubbard-Stratonovich transformation.
To do that, we Fourier transform to the momentum representation:
      $$
      \int d^3x \left( \bar\Psi(x) \sigma_3 \Psi(x) \right)^2 =
      \int \frac{dk_1}{(2\pi)^3}\frac{dk_2}{(2\pi)^3}\frac{dk_3}{(2\pi)^3}
      \bar \Psi(k_1)\sigma_3\Psi(k_2)
      \bar \Psi(k_3)\sigma_3\Psi(-k_1-k_2-k_3) \;.
      $$
There exist three independent ways of pairing two fast single-particle
momenta to form a slow two-particle momentum $q$: 
      \begin{eqnarray*}
        \begin{array}{ccccc}
          &\bar\Psi(k_1) &\Psi(k_2) &\bar\Psi(k_3) &\Psi(-k_1-k_2-k_3)\\ 
          a) &k &-k+q &k' &-k'-q \\ 
          b) &k &-k'-q &-k+q & k' \\
          c) &k &k' &-k'-q & -k+q
        \end{array}
      \end{eqnarray*}
Term $a)$ can be decoupled trivially, producing no more than energy 
shifts that can be absorbed by a redefinition of the chemical
potential.  The other two terms can be rearranged in the following
way.  For term $b)$ we have
      \begin{eqnarray*}
        && \sum_{k,k',q} \bar\Psi(k)\sigma_3\Psi(-k'-q)
        \bar\Psi(-k+q)\sigma_3\Psi(k')\\ \nonumber
        &=&\sum_{k,k',q} \bar\Psi(k)\sigma_3\Psi(-k'-q)
        \Psi^T(k')\sigma_3 \bar\Psi^T(-k+q)
        \nonumber \\
        &=& \sum_{k,k',q} \bar\Psi(k)\sigma_3\Psi(-k'-q)
          \left( -\bar\Psi(k')\gamma^{-1}{\cal C}\right) \sigma_3
          \left(\gamma {\cal C}^{-1}\Psi(-k+q)\right) \\
        &=& \sum_q {\rm STr}
        \left( \sum_{k'} \Psi(-k'-q)\bar \Psi(k') \sigma_3 \right)
        \left( \sum_{k}\Psi(-k+q)\bar\Psi(k)\sigma_3 \right) \;.
      \end{eqnarray*}
Here we have used ${\cal C}\sigma_3{\cal C}^{-1} = - \sigma_3$
and the relations
      $$
      \bar\Psi^T = \gamma {\cal C}^{-1}\Psi \;,
      \quad
      \Psi^T = -\bar\Psi\gamma^{-1} {\cal C} \;,
      $$
which follow from $\Psi = {\cal C}\gamma^{-1}\bar\Psi^T$ and $\bar\Psi
= -\Psi^T {\cal C}^{-1}\gamma$ (see the main text).  The term $c)$ is
the ``nice'' one and is easily put in the very same form, by using
only the cyclic invariance of the supertrace.  Thus, the quartic
interaction $(v^2/2) \int d^3x \; (\bar\Psi \sigma_3 \Psi)^2$ is
replaced by
      $$
      v^2 \sum_{|q| < q_0} \str \zeta(-q) \zeta(q) \;,
      $$
where $\zeta$ is given by a sum of dyadic products of the fields
$\Psi$ and $\bar\Psi$:
      $$
      \zeta(q) = \sum_{k} \Psi(-k+q)\bar \Psi(k) \sigma_3 \;.
      $$

\section{Effective action}
\label{sec:app_two}

We want to extract from the functional $\CS$ defined in (\ref{action})
a ``low-energy'' effective action for slowly varying fields $Q(x) =
T(x) Q_0 T^{-1}(x)$ generated from the saddle point $Q_0 = \sigma_3
\otimes \Sigma_3$.  To do that, we insert $Q(x) = T(x) Q_0 T^{-1}(x)$
into $\CS$:
      $$
      \CS[T Q_0 T^{-1}] = {1\over 2} \int d^3x \; \str
      \ln \left( {\cal H}_0 - \omega + {i\sigma_3\over 2\tau} T Q_0
      T^{-1} \right)(x,x) \;,
      $$
and expand with respect to $\omega$, $\tilde\Delta$, and $A$, and
up to two gradients of $Q(x)$, which are assumed to be ``small''
quantities.

Let us first extract the gradients.  For that, we set $\omega$,
$\tilde\Delta$ and $A$ to zero and consider
      $$
      S_0 = {1\over 2} \int d^3x \; \str
      \ln \left( \sigma_3 (
%{\nabla}
-{\nabla}^2/2m-\mu) + {i\sigma_3\over 2\tau} T Q_0
      T^{-1} \right)(x,x) \;.
      $$
This cannot readily be expanded as it stands, as $T(x)$ is ambiguous
by multiplication on the right by an arbitrary field $h(x)$ with $h(x)
Q_0 h^{-1}(x) = Q_0$.  To fix the ambiguity, we write $T(x)$ as
      $$
      T(x) = g {\rm e}^{X(x)} \quad {\rm where} \quad
      X(x) Q_0 + Q_0 X(x) = 0 \;.
      $$
The matrix $g$ is chosen in such a way that $g Q_0 g^{-1}$ coincides
with the local ``mean'' of $Q(x)$, while $X(x)$ parametrizes small
deviations from that mean.  The space dependence of $g$ is assumed to
be so weak that it can be neglected for the present purposes.

When $Q = g {\rm e}^X Q_0 {\rm e}^{-X} g^{-1}$ is inserted into $S_0$, the
constant matrix $g$ drops out by the cyclic invariance of the
supertrace, and expansion of the logarithm in powers of the small
quantity $X(x)$ gives
      \begin{eqnarray*}
        S_0 &=& {1 \over 2} \int d^3x \; \str \ln \left( {
%{\nabla}
-{\nabla}^2\over 2m} 
          - \mu + {i\over 2\tau} {\rm e}^X Q_0 {\rm e}^{-X} \right)(x,x) \\
        &=& - {1 \over 8\tau^2} \int d^3x \int d^3x' \;
        \str \left( {
%{\nabla}
-{\nabla}^2\over 2m} - \mu + {i\over 2\tau}
          Q_0 \right)^{-1}(x,x') \left( X(x')-X(x) \right) \\
          &&\hspace{3.5cm} \times \left( {
%{\nabla}
-{\nabla}^2\over 2m} - \mu + 
            {i\over 2\tau} Q_0 \right)^{-1}(x',x) 
        \left( X(x)-X(x') \right) + ... \;.
      \end{eqnarray*}
We assume the elastic mean-free path $\ell$ to be much larger than
the Fermi wave length $(k_F \ell \gg 1)$.  The Green functions then
are well approximated by
      $$
      \left( {
%{\nabla}
-{\nabla}^2\over 2m} - \mu \pm {i \over 2\tau} \right)^{-1}(x,x')
      \simeq {m \over 2\pi} {\exp (\mp ik_F |x-x'|) \over |x-x'|}
      \; {\rm e}^{-|x-x'|/2\ell} \;,
      $$
and Taylor expansion of the fields
      $$
      X(x) - X(x') = \sum_{k=1}^3 (x^k - {x'}^k) 
      {\partial X \over \partial x^k} + ...
      $$
leads to the integral 
      $$
      \int d^3x' \; {|x - x'|^2 \over 3} \left| \left( {
%{\nabla}
-{\nabla}^2\over 2m} 
        - \mu + {i \over 2\tau} \right)^{-1} (x,x') \right|^2 =
      {2m^2 \ell^3 \over 3\pi} \;.
      $$
Assembling factors we obtain
      $$
      S_0 = {m^2 \ell^3 \over 12 \pi \tau^2} \int d^3x \;
      \str (\nabla X)^2 \;.
      $$
By recalling the formulas $\nu = m k_F / 2\pi^2 = m^2 \ell / 2\pi^2
\tau$ for the local density of states and $D = \ell^2 / 3\tau$ for
the diffusion constant in three dimensions, we can write this as
      $$
      S_0 = {\pi\nu D \over 2} \int d^3x \; \str (\nabla X)^2 \;.
      $$
To put the result in invariant form, we use
      $$
      \str (\nabla Q)^2 = 
      \str \left( \nabla (g {\rm e}^X Q_0 {\rm e}^{-X} g^{-1})
        \right)^2 = \str (\nabla {\rm e}^{2X})(\nabla {\rm e}^{-2X}) 
        = - 4 \str (\nabla X)^2 + ... \;.
      $$
This gives
      $$
      S_0 = - {\pi\nu D \over 8} \int d^3x \; \str (\nabla Q)^2 \;.
      $$
which is the kinetic part of the effective action (\ref{fseff}).

The dependence of the Lagrangian on the magnetic field is easily
restored by the following symmetry argument.  Consider a gauge
transformation
      $$
      A \mapsto A + {1 \over e} \nabla \varphi \;, \quad
      \psi \mapsto {\rm e}^{i\varphi\sigma_3} \psi \;.
      $$
The transformation law induced on the composite field $Q \sim 
\Psi\bar\Psi\sigma_3$ is
      $$
      Q \mapsto {\rm e}^{i\varphi\sigma_3} Q {\rm e}^{-i\varphi\sigma_3} \;,
      $$
so the gradient $\nabla Q$ transforms as
      $$
      \nabla Q \mapsto \nabla Q + i(\nabla\varphi) [\sigma_3 , Q ] \;.
      $$
This determines the covariant derivative $\nabla_A$ acting on $Q$ to
be
      $$
      \nabla_A = \nabla - i e A [\sigma_3 , \cdot ] \;.
      $$
Therefore, the unique way of making the effective action gauge invariant is to 
substitute
      $$
      \str (\nabla Q)^2 \to \str (\nabla_A Q)^2 \;,
      $$
which restores the dependence on the magnetic field without any
calculation.

Finally, we expand (\ref{action}) to extract the terms linear in
$\tilde\Delta$ and $\omega$.  These are
      $$
      S_1 = {1 \over 2} \int d^3x \; \str 
      \left( {-{\nabla}^2\over 2m}
        -\mu + {i \over 2\tau}Q \right)^{-1}(x,x) \; \sigma_3 
      (\tilde\Delta - \omega) \;.
      $$
By using the saddle point equation $(
%{\nabla}
-{\nabla}^2/2m - \mu + iQ/2\tau)^{-1}
(x,x) = - i \pi \nu Q(x)$ and the relation $\sigma_3 \tilde\Delta = -
\tilde\Delta \sigma_3$ we immediately get
      $$
      S_1 = {i\pi\nu \over 2} \int d^3x \; \str Q (\tilde\Delta 
      + \omega) \sigma_3 \;.
      $$
(Note that, since $Q$ and $\tilde\Delta$ behave in the same way under
gauge transformations, the term $\str Q\tilde\Delta\sigma_3$ is gauge
invariant, as is $\str Q\omega\sigma_3$.)  Summing the terms $S_0$ and
$S_1$ we arrive at the effective action (\ref{fseff}).  The sign
differences compared to \cite{ats} come from: a conjugation of $Q$
by $\sigma_3$ (which changes the sign of the off-diagonal terms of $Q$
in the particle-hole space), and an overall change of sign of $Q$.

\section{Differential equation}
\label{sec:app_three}

To solve the Usadel equation for a vortex configuration, we made the
ansatz $Q_0 = g(r) \sig_3 \otimes \Sig_3 + f(r) \sig_2 {\rm e}^{
  -i\varphi \sig_3}\otimes 1_{cc}$ in Section \ref{subsec:ssp}.  The
purpose of this appendix is to derive the differential equation obeyed
by the coefficient functions $f$ and $g$.

To begin, we recall the Usadel equation for $Q$:
      $$
      D \nabla_A (Q\nabla_A Q) = - i [Q , (\tilde\De+\om)\sig_3 ] \;.
      $$
Since all of the quantities $Q_0$, $A\sig_3$, $\tilde\Delta = |\Delta|
\sig_1 \exp(-i\varphi\sig_3)$ and $\om = E \Sigma_3$ are diagonal 
in charge conjugation space, we may specialize to the sector where
$\Sigma_3$ equals unity and work with the simpler ansatz
      $$
      Q = g(r) \sig_3 + f(r) \sig_2 {\rm e}^{-i\varphi\sig_3} \;.
      $$
(Alternatively, we may imagine absorbing $\Sigma_3$ into $g$.)  The
right-hand side of the Usadel equation then becomes
      $$
      -i [ g \sig_3 , \tilde\Delta\sig_3]
      -i [ f \sig_2 {\rm e}^{-i\varphi\sig_3}, E \sig_3]
      = \left( i |\Delta| g + E f \right)
      2 \sig_1 {\rm e}^{-i\varphi\sig_3} \;.
      $$

The evaluation of the left-hand side of the Usadel equation is more
laborious.  From $\nabla_A = \nabla - ie A [\sig_3 , \cdot]$ we have
      $$
      \nabla_A (Q\nabla_A Q) = 
      \nabla (Q \nabla Q)
      - ie \nabla ( Q [A\sig_3,Q] )
      - ie A [ \sig_3, Q\nabla Q] 
      - e^2 A^2 [ \sig_3 , Q [\sig_3,Q] ] \;.
      $$
We now express this in cylindrical coordinates $(r,\vphi,z)$.  For the
term with two derivatives we get
      \begin{eqnarray*}
        \nabla (Q \nabla Q) &=& 
        {1 \over r} \partial_r (r Q \partial_r Q)
        + {1 \over r^2} \partial_\varphi (Q \partial_\varphi Q)
        + \partial_z (Q \partial_z Q) \\
        &=& {1 \over r} \partial_r (r f \partial_r f + r g \partial_r g)
        + {i \over r} \partial_r (r f \partial_r g - r g \partial_r f)
        \sig_1 {\rm e}^{-i\varphi \sig_3} \\
        &+& {1 \over r^2} \partial_\varphi \left(
          (g\sig_3 + f \sig_2 {\rm e}^{-i\varphi\sig_3})
          f {\rm e}^{i\varphi \sig_3} \sig_1 \right) \\
        &=& \left( 
          {1 \over r} \partial_r (r f \partial_r g - r g \partial_r f)
          + {fg \over r^2} \right) i\sig_1 {\rm e}^{-i\varphi\sig_3} \;.
      \end{eqnarray*}
The term $f\partial_r f + g\partial_r g$ vanishes by $f^2 + g^2 = 1$.
Taking the magnetic vector potential to be purely azimuthal, $A = 
e_\varphi A_\varphi(r)$, we obtain for the part linear in $A$
      \begin{eqnarray*}
        &-& ie \nabla (Q [A\sig_3,Q]) - ie [A\sig_3, Q\nabla Q] =
        - {ie \over r} \partial_\varphi (Q[A_\varphi\sig_3,Q]) 
        - {ie \over r} [A_\vphi\sig_3, Q\partial_\varphi Q] \\
        &=&
        - {2eA_\varphi \over r}\partial_\vphi \left( (g\sig_3 + f \sig_2 
          {\rm e}^{-i\varphi \sig_3}) f {\rm e}^{i\varphi \sig_3}
          \sig_1 \right)
        - {ieA_\varphi \over r} \left[ \sig_3 , (g \sig_3 + f \sig_2 
          {\rm e}^{-i\varphi \sig_3}) f {\rm e}^{i\varphi \sig_3}
        \sig_1 \right] \\
      &=& - 4eA_\varphi {fg \over r} \; i \sig_1 {\rm e}^{-i\varphi\sig_3} \;.
    \end{eqnarray*}
The term quadratic in the vector potential gives
      $$
      - e^2 A^2 \left[\sig_3,Q[\sig_3,Q]\right] = 
      - e^2A_\varphi^2 (\sig_3 Q \sig_3 Q - Q\sig_3 Q \sig_3)
      = 4 e^2 A_\vphi^2 f g \; i \sig_1{\rm e}^{-i\varphi \sig_3} \;.
      $$
Gathering all terms and dividing by $i\sig_1{\rm e}^{-i\varphi
  \sig_3}$ we arrive at the equation
      $$
        {1 \over r}\partial_r (r f\partial_r g - r g\partial_r f)
        + \left( {1 \over r} - 2eA_\varphi \right)^2 f g
        = {2 \over D} \left( |\Delta| g - i E f \right) \;,
      $$
whose solution for $E = 0$ is discussed in the text.

\section{Choosing between solutions}
\label{sec:app_four}

The full saddle point equation (including $\De$ and $\om$) is
difficult to solve exactly if the gap function varies in space.
However, for the case of a uniform gap a solution is easy to come by.
This solution is constant and describes the region far from the
vortex.  For values of the quasiparticle energy $\om$ large compared
to $\De_0$, the solution should approach the normal metal solution.
This fact will now be used to analytically continue the metallic
solution ($\om/\De_0 \gg 1$) to the zero-energy solution for a
superconductor ($\om = 0$).

\begin{figure}
        \hspace{2.0cm} \epsfxsize=10cm \epsfbox{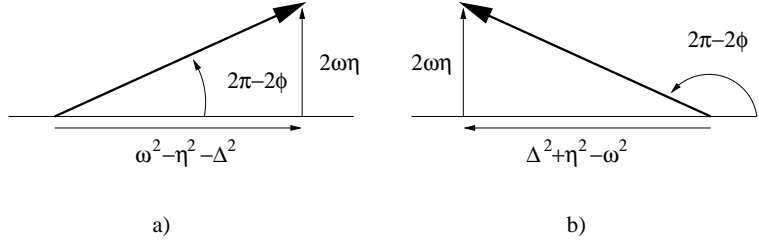}
        \caption{Definition of the angle $\phi$ for a) 
          $\om^2-\eta^2-\Delta^2>0$ and b) $\om^2-\eta^2-\Delta^2<0$.}
        \label{fig:angle}
      \end{figure}

If we ignore the charge conjugation degree of freedom, which is not
essential here, the full saddle point equation reads: 
      $$
      Q(x) = \frac{i}{\pi\nu} \left\langle x \left\vert
      \frac{1}{h_0+ iQ/2\tau + \sig_3 \tilde
      \De-(\om-i\eta)\sig_3}\right\vert x \right\rangle \;.  
      $$
In the absence of any spatial dependence in the gap $\De$, this is
solved by Fourier transforming to momentum space and taking $Q$ to be
constant:
      $$
      Q = \frac{i}{\pi \nu} \int {{d}^3 k \over (2\pi)^3}
      \; \frac{1}{k^2/2m - \mu + iQ/2\tau + \sig_3 \tilde\De - 
      (\om-i\eta)\sig_3} \;.  
      $$
Here, $\om > 0$ and $\eta > 0$.  We choose the gauge where $\tilde\De
= \De \sig_1$ and diagonalize the matrix $ M = i \De \sig_2 -(\om -
i\eta)\sig_3$ by a similarity transformation with a matrix $T$:
      $$
        TMT^{-1} = \( 
        \begin{array}{cc}
           -\sqrt{(\om-i\eta)^2-\De^2} & 0 \\
           0 & \sqrt{(\om-i\eta)^2-\De^2} 
        \end{array}\)=\la \sig_3 \;.
      $$
The sign of the first eigenvalue, $\lambda$, of $M$ can be chosen so
that $\lambda \to -\om+i\eta$ when $\om/\De \to \infty$. We then
have:
      $$
      \la=\( (\om^2-\eta^2-\De^2)^2+4\eta^2\om^2 \)^{1/4} e^{i \phi}\;,
      $$
where the angle $\phi$ varies from $\pi$ to $\pi/2$ as $\om$ goes from
infinity to $0$ (see Figure \ref{fig:angle}).  The imaginary part of
$\la$ remains positive for any (positive) value of $\om$.

The matrix that diagonalizes $M$ is $T=(1+b\sig_1)/\sqrt{1-b^2}$, with
$b=-\frac{\om-i\eta}{\De}+\sqrt{\(\frac{\om-i\eta}{\De}\)^2-1}$.  It
is important that the imaginary part of $b$ is negative for any value
of $\om/\De$.  Note in particular that for $\om = 0$ we have $b = -i$.
The saddle point equation now takes the form:
      $$
      \bar Q = 
      \frac{i}{\pi \nu}\int {{d}^3 k \over (2\pi)^3} 
      \; \frac{1}{k^2/2m - \mu + i\bar Q/2\tau + \la\sig_3} \;.
      $$
The physical solution is given by $\bar Q = TQT^{-1} = {\rm sign}({\rm
Im} \la) \times \sig_3 = \sig_3$.  Therefore
      $$
      Q = T^{-1} \sig_3 T =
      \frac{1+b^2}{1-b^2}\sig_3+\frac{2ib}{1-b^2}\sig_2 \;.
      $$
Unphysical solutions (which break causality) also exist. 

By comparing to the ansatz used in Section \ref{subsec:ssp} to solve
the Usadel equation, $Q = \sig_3 \cos \th + \sig_2 \sin\th$, we
conclude that $b = -i \tan\frac{\th}{2}$.  In particular, for $\om=0$
where $b = -i$, we get $\tan\frac{\th}{2} = 1$.  Therefore, $\th =
\frac{\pi}{2} ({\rm mod} 2\pi)$ is the zero-energy solution for the
homogeneous superconductor, in agreement with the asymptotics found in
Section \ref{subsec:ssp} from a different argument.

\section{Vanishing vertices}
\label{sec:app_five}

In this appendix we show that the terms on the second line in equation
\ref{fourthorder} do not contribute to the transmittance $\tau(E)$ in
one-loop order.  Let us consider the first and second term, which give
      $$
      \left\langle \partial Z_0^{B1,B2}(0) 
        \partial\tilde Z_0^{B2,B1}(1) \; 
        \str Z_0 \partial\tilde Z_0 
        (Z_1 \partial\tilde Z_1 - \partial Z_1 \tilde Z_1) \Big|_s
      \right\rangle_0 \;.
      $$
As before, only the contractions of type $1$ and $2$ in
Fig.~\ref{fig:diagram} are nonzero, and we obtain
      $$
      {\rm D}_0(\dot 0,\dot s)
      \( {\rm D}_\wedge(\dot s,s) - {\rm D}_\wedge(s,\dot s) \) 
      {\rm D}_0(s,\dot 1) + (\wedge \to \vee) \;,
      $$
where the dot over an argument represents a derivative acting on it.
Because the first derivatives of the propagators ${\rm D}_0(s,s')$,
${\rm D}_\wedge(s,s')$ and ${\rm D}_\vee(s,s')$ are discontinuous at
$s = s'$, these terms are ambiguous and have to be defined by some
point-splitting procedure, {\it e.g.}
      $$
      {\rm D}_\wedge (\dot s,s) = \lim_{a\to 0}\frac{1}{a}
      \( {\rm D}_\wedge(s+\frac{a}{2},s)-{\rm D}_\wedge(s-\frac{a}{2},s)\) 
      = \frac{\sinh \sqrt{i\ep}(1-2s)}{2 \sinh{\sqrt{i\ep}}} \;.
      $$
Since the propagators are symmetric in their arguments, we have
      $$
      {\rm D}_{\wedge,\vee}(\dot s, s)={\rm D}_{\wedge,\vee}(s,\dot s)\;,
      $$
which establishes our claim.  In the same way it is seen that the
one-loop contributions from the third and fourth term in equation
\ref{fourthorder} cancel each other.

\section{Invariant integration measure}
\label{sec:app_six} 

The field space of the nonlinear $\sigma$ model we have derived is a
coset space $G/H$ with $G = {\rm OSp}(2n|2n)$ and $H = {\rm GL}
(n|n)$. The space $G/H$ is realized by matrices
      $$
      q = g \Sigma_3 g^{-1} \;,
      $$
where $g\in G$, and $\Sigma_3$ is fixed by the elements $h$ of the
subgroup $H$:
      $$
      h \Sigma_3 h^{-1} = \Sigma_3 \;.
      $$
It is easily seen that the symmetry relation $q = -\gamma q^T 
\gamma^{-1}$ corresponds to the condition
      $$
      g = \gamma (g^{-1})^T \gamma^{-1} \;,
      $$
which defines the group $G$. Let $D\mu(q)$ denote the superintegration 
measure (on $G/H$) invariant under $q\mapsto g_0^{\phantom{1}} q
g_0^{-1}$ for all $g_0\in G$ and normalized by $\int D\mu(q) = 1$. Our
goal in this appendix is to express $D\mu(q)$ in the parametrization
of $q$ introduced in Section \ref{sec:q}.

Quite generally, if we are given a parametrization with
(super)coordinates $\xi^i$ and metric tensor
      $$
      {\rm STr}\, ({\rm d}q)^2 = 
      \sum_{ij}{\rm d}\xi^i g_{ij}(\xi ){\rm d}\xi^j\,,
      $$
then the invariant integration measure is locally ({\it i.e.} modulo
boundary ambiguities) expressed by
      $$
      D\mu(q) = D\xi \; {\rm SDet}^{1/2}\big( g_{ij}(\xi )\big) \;,
      $$
where $D\xi$ is a (suitably normalized) Euclidean measure. Doing the 
computation this way requires knowledge of all matrix elements of the
metric tensor and is more complicated than is necessary.

Since we are parametrizing $q$ in the form 
      $$
      q(\xi) = g(\xi )\Sigma_3 g^{-1}(\xi ) \,,
      $$
a more efficient approach is to calculate the measure directly from 
the Jacobian of the map
      $$
      \left(\xi^1,\ldots ,\xi^{\rm dim}\right) =\xi\mapsto
      g(\xi )\Sigma_3 g^{-1}(\xi ) \;.
      $$
However, this is still not the best method as it involves the group
element $g$ {\it twice}, once on the left and once on the right (by
the inverse).

The most efficient approach is to turn to the ``coset space picture''.
Given some basis $X_i,\ldots ,X_{\rm dim}$ of ${\cal P} \stackrel{
  \sim}{=} T_H(G/H)$, we assemble the coordinate functions $\xi^i$
into a matrix--valued function $X\equiv\sum_i\xi^iX_i\,:\; G/H\to
{\cal P}$ and consider some map
      $$
      X\mapsto g(X)H\,,
      $$
which sends $X$ into the coset space $G/H$.  Its differential is 
      $$
      (X,{\rm d}X) \mapsto \frac{d}{ds}
      \Big|_{s=0}g(X+s\cdot{\rm d}X)H\;.
      $$
Because the integration measure on $G/H$ we wish to express by $X$,
is invariant under left translations, we may multiply by $g(X)^{-1}$
and switch to a map $T_X: {\cal P}\to {\cal P}$ defined by
      \begin{eqnarray*}
        T_X({\rm d}X) & = & \frac{d}{ds}\bigg|_{s=0}g(X)^{-1}
        g(X+s\cdot{\rm d}X)H\\ 
        & = & \left( \frac{d}{ds}\bigg|_{s=0}
          g(X)^{-1}g(X+s\cdot{\rm d}X) \right)_{\cal P}\;.
      \end{eqnarray*}
The subscript ${\cal P}$ means projection of a Lie algebra element $A
\in {\rm Lie}(G)={\rm Lie}(H)+{\cal P}$ onto $A_{\cal P}\in {\cal P}$.
Let $J(X)={\rm SDet}\, T_X$ be the Berezinian (or superjacobian) of
the transformation. Then the invariant measure $D\mu (q)$ is simply
expressed by
      $$
      DX\,J(X)\;.
      $$

We now wish to apply this general formalism to the parametrization
      $$
      q = t_0^{\phantom{1}}t_1^{\phantom{1}}\Sigma_3^{\phantom{1}}
      t_1^{-1}t_0^{-1} \;, \quad
      t_{\alpha} = \pmatrix{1 & Z_{\alpha}\cr
        \tilde Z_{\alpha} & 1\cr} \quad
      (\alpha =0,1) \;.
      $$
This is not possible immediately, as the matrices $t_{\alpha}$
do not obey the law $g = \gamma (g^{-1})^T \gamma^{-1}$
defining the group $G = {\rm OSp}(2n|2n)$ (although $q=t_0^{\phantom{1}}
t_1^{\phantom{1}}\Sigma_3^{\phantom{1}}t_1^{-1}t_0^{-1}$ does satisfy 
$q=-\gamma q^T\gamma^{-1}$). To be on safe mathematical ground, 
it is advisable to start from a parametrization
      $$
      q = \tau_0^{\phantom{1}}\tau_1^{\phantom{1}}\Sigma_3^{\phantom{1}}
      \tau_1^{-1}\tau_0^{-1}
      $$
such that the $\tau_{\alpha}$ do satisfy the group law. This is achieved by 
choosing
      \begin{eqnarray*}
        \tau_{\alpha} & = & \pmatrix{
          (1-Z_{\alpha}\tilde Z_{\alpha})^{-1/2} &
          Z_{\alpha}(1-\tilde Z_{\alpha}Z_{\alpha})^{-1/2}\cr
          \tilde Z_{\alpha}(1-Z_{\alpha}\tilde Z_{\alpha})^{-1/2} &
          (1-\tilde Z_{\alpha}Z_{\alpha})^{-1/2}} \\
        & = & \pmatrix{ 1 & Z_{\alpha}\cr \tilde Z_{\alpha} & 1\cr} 
        \pmatrix{h^+_{\alpha} & 0\cr 0 & h^-_{\alpha} \cr}
        = t_{\alpha}h_{\alpha} \;,
      \end{eqnarray*}
where $h^+_{\alpha}=(1-Z_{\alpha}\tilde Z_{\alpha})^{-1/2}$
and $h^-_{\alpha} =(1-\tilde Z_{\alpha}Z_{\alpha})^{-1/2}$.

The general formalism now tells us 
      $$
      D\mu (q) = D(Z_0,\tilde Z_0)\,D(Z_1,\tilde Z_1)\,J(Z_0,\tilde Z_0,
      Z_1,\tilde Z_1) \;,
      $$
where $D(Z_0,\tilde Z_0)$ and $D(Z_1,\tilde Z_1)$ are flat Berezin 
measures, $J={\rm SDet}\,T$ is the superjacobian, and 
      $$
        T({\rm d}Z_0,\ldots ) = {\Big( \tau_1^{-1}\tau_0^{-1}
          {\rm d}(\tau_0^{\phantom{1}}\tau_1^{\phantom{1}})
          \Big)}_{\cal P} \\
        = {\Big(\tau_1^{-1}{\rm d}\tau_1^{\phantom{1}}\Big)}_{\cal P}
        +{\Big(\tau^{-1}_1(\tau_0^{-1}{\rm d}\tau_0^{\phantom{1}})
          \tau_1^{\phantom{1}}\Big)}_{\cal P} \;.
      $$
Recall the decomposition $Z = Z_0 + Z_1$ which, in terms of spaces, we
write ${\cal P}={\cal P}_0+{\cal P}_1$.  Because of ${(\tau_1^{-1}{\rm
    d}\tau_1^{\phantom{1}})}_{\cal P}= {(\tau_1^{-1} {\rm d} \tau_1^{
    \phantom{1}})}_{{\cal P}_1}$ and the general relation 
      $$
      {\rm SDet} \pmatrix{X & Y\cr 0 & Z\cr} = 
      {\rm SDet}\pmatrix{X & 0\cr 0 & Z\cr} \;,
      $$
we have ${\rm SDet}\,T={\rm SDet}\,T'$ with
      $$
      T'({\rm d}Z_0,\ldots )={\Big(\tau_1^{-1}{\rm d}\tau_1^{\phantom{1}}
        \Big)}_{{\cal P}_1}
      +{\Big( \tau_1^{-1}(\tau_0^{-1}{\rm d}\tau_0^{\phantom{1}})
        \tau_1^{\phantom{1}}\Big)}_{{\cal P}_0}\;.
      $$
If $X$ is any element of ${\rm Lie}(G)$ we introduce the notation
      $$
      X = \pmatrix{A(X) & B(X)\cr C(X) & D(X) \cr} \;,
      $$
where $\pmatrix{A & 0\cr 0 & D\cr} \in {\rm Lie}(H)$, and $\pmatrix{
  0&B\cr C & 0\cr} \in {\cal P}$.  A straightforward calculation then
gives
      \begin{eqnarray*}
        B(\tau_1^{-1}{\rm d}\tau_1^{\phantom{1}}) & = &
        h_1^+{\rm d}Z_1^{\phantom{1}}h_1^-\,\\
        C(\tau_1^{-1}{\rm d}\tau_1^{\phantom{1}}) & = & 
        h_1^-{\rm d}\tilde Z_1^{\phantom{1}}h_1^+\,, 
        \qquad {\rm and}
      \end{eqnarray*}
      \begin{eqnarray*}
        B\Big( {(\tau_1^{-1}\tau_0^{-1}{\rm d}\tau_0^{\phantom{1}}
          \tau_1^{\phantom{1}})}_{{\cal P}_0}\Big) & = &
        h_1^+\Big( h_0^+{\rm d}Z_0^{\phantom{1}}h_0^-
        -Z_1^{\phantom{1}}h_0^-{\rm d}\tilde Z_0^{\phantom{1}}h_0^+
        Z_1^{\phantom{1}}\Big) h_1^-\,,\\
        C\Big( {(\tau_1^{-1}\tau_0^{-1}{\rm d}\tau_0^{\phantom{1}}
          \tau_1^{\phantom{1}})}_{{\cal P}_0}\Big) & = &
        h_1^-\Big( h_0^-{\rm d}\tilde Z_0^{\phantom{1}}h_0^+
        -\tilde Z_1^{\phantom{1}}h_0^+{\rm d}Z_0^{\phantom{1}}h_0^-
        \tilde Z_1^{\phantom{1}}\Big) h_1^+\,.
      \end{eqnarray*}
By writing out the matrices in explicit form, one easily sees that
the Berezinian of the transformation ${\rm d}Z_1^{\phantom{1}} \mapsto
h_1^+{\rm d}Z_1^{\phantom{1}}h_1^-,\; {\rm d}\tilde Z_1^{\phantom{1}}
\mapsto h_1^-{\rm d}\tilde Z_1^{\phantom{1}}h_1^+$ is given by ${\rm
  SDet}(h_1^+)\, {\rm SDet}(h_1^-)={\rm SDet}(1-\tilde Z_1^{\phantom{
    1}} Z_1^{\phantom{1}})$, while the Berezinian of the
transformation ${\rm d}Z_0^{\phantom{1}}\mapsto h_0^+{\rm
  d}Z_0^{\phantom{1}}h_0^-,\; {\rm d}\tilde Z_0^{\phantom{1}}\mapsto
h_0^-{\rm d}\tilde Z_0^{\phantom{1}}h_0^+$ is simply unity as a result
of cancellations due to supersymmetry.

For similar reasons, the Berezinian of the transformation
      \begin{eqnarray*}
        {\rm d}Z_0^{'} & \mapsto & h_1^+({\rm d}Z_0^{'} -Z_1^{\phantom{1}}
        {\rm d}{\tilde Z}_0^{'}
        Z_1^{\phantom{1}})h_1^-\,,\\
        {\rm d}{\tilde Z}_0^{'} & \mapsto & 
        h_1^-({\rm d}{\tilde Z}_0^{'} -{\tilde Z}_1^{\phantom{1}}
        {\rm d}Z_0^{'}{\tilde Z}_1^{\phantom{1}})h_1^+
      \end{eqnarray*}
turns out to be unity. Thus, by collecting factors the invariant
integration measure $D\mu(q)$ in the parametrization
$q = \tau_0^{\phantom{1}}\tau_1^{\phantom{1}}\Sigma_3^{\phantom{1}}
\tau_1^{-1}\tau_0^{-1}$ is given by
      \begin{equation}
        D(Z_0,\tilde Z_0)\,D(Z_1,\tilde Z_1)\,{\rm SDet}(1-\tilde Z_1Z_1)\;.
        \label{Dmuq}
      \end{equation}
Finally, we switch to the desired parametrization
      $$
      Q = t_0^{\phantom{1}}t_1^{\phantom{1}}\Sigma_3^{\phantom{1}}
      t_1^{-1}t_0^{-1}\\
      = t_0^{\phantom{1}}\tau_1^{\phantom{1}}\Sigma_3^{\phantom{1}}
      \tau_1^{-1}t_0^{-1}\\
      = \tau_0^{\phantom{1}}h_0^{\phantom{1}}\tau_1^{\phantom{1}}
      \Sigma_3^{\phantom{1}}\tau_1^{-1}h_0^{-1}\tau_0^{-1}\;.
      $$
It is seen that this amounts to transforming 
$\tau_1^{\phantom{1}}\mapsto h_0^{\phantom{1}}\tau_1^{\phantom{1}}h_0^{-1}$
or, equivalently,
      $$
      Z_1^{\phantom{1}}\mapsto h_0^+Z_1^{\phantom{1}}h_0^-,\quad
      \tilde Z_1^{\phantom{1}}\mapsto h_0^-
      \tilde Z_1^{\phantom{1}}h_0^+\;.
      $$
Once again, the superjacobian of this transformation is unity, so that
$D\mu (q)$ in the desired parametrization has the form (\ref{Dmuq})
as it stands.

In one-loop order of the perturbation expansion for the
one-dimensional field theory, the factor ${\rm SDet}(1 - \tilde Z_1
Z_1)$ contributes a two-vertex $a^{-1} \int {d}s \; \str \tilde
Z_1 Z_1$ where $a$ is the regularization scale.  This vertex does not
couple to the external legs $\partial Z_0^{B1,B2}$ and $\partial\tilde
Z_0^{B2,B1}$.  Therefore we may pretend the functional integration
measure to be flat, which is what we did.

\end{document}